\documentclass{article}
\usepackage{amsmath,amsthm,amssymb}
\usepackage[a4paper]{geometry}
\usepackage{graphicx}
\usepackage[textwidth=8em,textsize=normalsize]{todonotes}
\usepackage{natbib}
\usepackage{lipsum}
\usepackage{float}
\usepackage{subcaption}
\usepackage{dcolumn}
\usepackage{url}
\usepackage{xcolor}
\usepackage{multirow}
\usepackage[linesnumbered,ruled]{algorithm2e}

\theoremstyle{plain}
\newtheorem{prop}{Proposition}

\newtheorem{lemm}[prop]{Lemma}
\newtheorem{theo}[prop]{Theorem}

\theoremstyle{definition}

\newtheorem{defi}{Definition}
\newtheorem{assu}{Assumption}
\newtheorem{rema}{Remark}

\DeclareMathAlphabet{\pazocal}{OMS}{zplm}{m}{n}

\def\Node{{\pazocal N}}
\def\XX{{\pazocal X}}

\usepackage{template_tex}

\newcommand\blfootnote[1]{%
  \begingroup
  \renewcommand\thefootnote{}\footnote{#1}%
  \addtocounter{footnote}{-1}%
  \endgroup
}

\begin{document}

\title{Estimating heterogeneous treatment effects with right-censored data via causal survival forests}
\author{Yifan Cui \blfootnote{Authors listed in alphabetical order} \\
Zhejiang University
\and
Michael R. Kosorok \\
UNC Chapel Hill
\and
Erik Sverdrup \\
Stanford University
\and
Stefan Wager \\
Stanford University
\and
Ruoqing Zhu \\
UIUC}

\maketitle

\begin{abstract}
Forest-based methods have recently gained in popularity for non-parametric treatment effect estimation.
Building on this line of work, we introduce causal survival forests, which can be used to estimate heterogeneous treatment
effects in a survival and observational setting where outcomes may be right-censored. Our approach relies on orthogonal
estimating equations to robustly adjust for both censoring and selection effects under unconfoundedness.
In our experiments, we find our approach to perform well relative to a number of baselines.
\end{abstract}

%\keywords{Precision medicine, Random forests, Right-censored data, Heterogeneous treatment effects, Confidence intervals}

\section{Introduction}

The problem of heterogeneous treatment effect estimation plays a central role in data-driven personalization
and, as such, has received considerable attention in the recent literature, including  \citet{athey2016recursive}, \citet{foster2019orthogonal},
\citet*{foster2011subgroup}, \citet*{hahn2017bayesian}, \citet*{kennedy2020optimal}, \citet*{kunzel2019metalearners},
\citet*{luedtke2016super},  \citet{nie2020quasi}, \citet*{tian2014interaction} and \citet*{wager2018estimation}.
One difficulty in applying this line of work to medical applications, however, is that such applications
often involve potentially censored survival outcomes---and existing methods for treatment heterogeneity often cannot be used in this setting.

To address this challenge, we propose {\it causal survival forests}, an adaptation of
the causal forest algorithm of \citet*{wager2019grf} that adjusts for censoring using doubly robust estimating
equations developed in the survival analysis literature \citep{tsiatis2007semiparametric,van2003unified}. The method is robust, computationally tractable, and  outperforms available baselines in our experiments. We also study the asymptotic behavior of causal survival forests,
and establish conditions under which its predictions are consistent and asymptotically normal.

We focus on a statistical setting where, for $i = 1, \, \ldots, \, n$ training examples, we
assume independent and identically distributed (i.i.d.) tuples $\cb{X_i, \, T_i, \, C_i, \, W_i}$ where $X_i \in \xx$ denote
covariates, $T_i \in \RR_+$ is the survival time for the $i$-th unit, $C_i \in \RR_+$ is the time at which the $i$-th
unit gets censored, and $W_i \in \cb{0, \, 1}$ denotes treatment assignment. The goal of a statistician is to measure the average effect of
the treatment $W_i$ on survival time $T_i$ conditional on $X_i = x$. We define causal effects in terms of the standard potential outcomes framework \citep{imbens2015causal}, i.e., we posit potential outcomes $\cb{T_i(0), \, T_i(1)}$
such that $T_i = T_i(W_i)$, and seek to estimate the conditional average treatment effect on the survival time,
\begin{equation}
\label{eq:cate}
\tau(x) = \EE{T_i(1) - T_i(0) \cond X_i = x},
\end{equation}
or some relevant transformation of the survival time.
However, unlike in the usual setting for heterogeneous treatment effect estimation discussed in, e.g.,
\citet{kunzel2019metalearners} and \citet{nie2020quasi}, we do not always get to see the realized survival time $T_i$;
rather, we only observe $U_i = T_i \land C_i$ along with a non-censoring indicator $\Delta_i = 1\cb{T_i \leq C_i}$.
The main challenge for causal survival forests is thus in estimating $\tau(x)$ in \eqref{eq:cate} while only
observing $U_i$ and $\Delta_i$ as opposed to the target outcome $T_i$.

\begin{figure}[t]
\centering
\begin{tabular}{c}
    \includegraphics[width=0.8\textwidth]{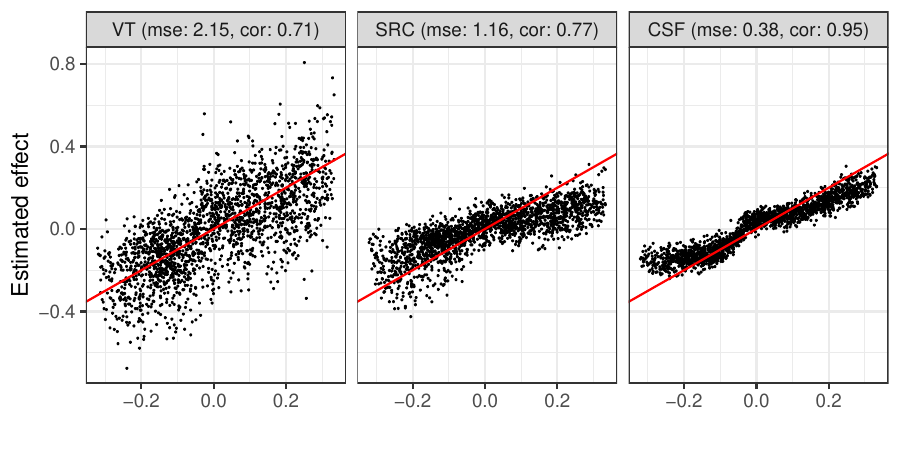} \\
    \includegraphics[width=0.8\textwidth]{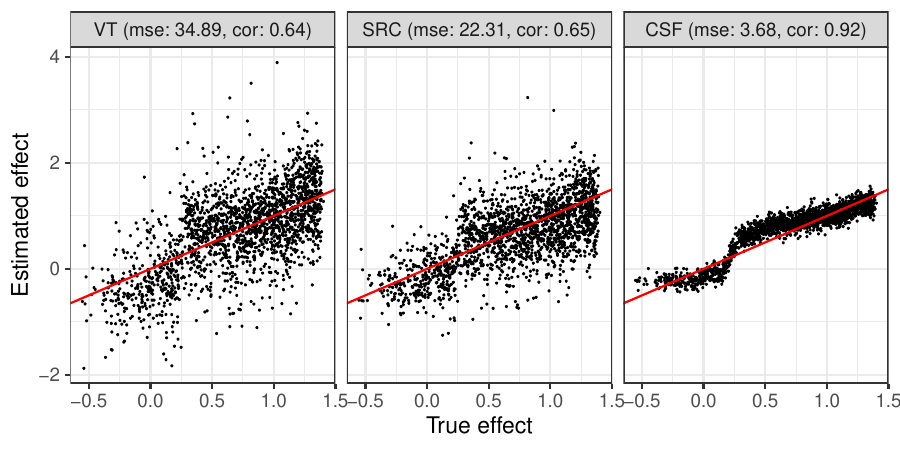}
\end{tabular}

\caption{Conditional average treatment effect (CATE) predictions versus ground truth for the proposed method (``CSF'') with two alternatives, ``VT'': virtual twins, ``SRC'': the $S$-learner. The top row is for a data generating process with continuous survival time (the second kind described in section \ref{sec:simulations}) and the bottom row for a data generating process with discrete response (the third kind described in section \ref{sec:simulations}). Each estimator is fit on a sample size of 5 000, with default package tuning parameters (detailed in Section \ref{sec:simulations}), predictions are for a test set of size 2 000. Mean squared error (MSE) and correlation with the true $\tau(x)$ is in parentheses. MSE is multiplied by 100 for readability.}

\label{fig:figure1}
\end{figure}

To illustrate the promise of causal survival forests, consider the following simple simulation experiment. We
compare our proposed method (CSF) to both an adaptation of the virtual twins method (VT) proposed by
\citet*{foster2011subgroup}, and a variant of what \citet{kunzel2019metalearners} call the $S$-learner approach implemented
using the survival forests from \citet*{ishwaran2008random} (SRC).
We defer the details of these implementations to Section \ref{sec:simulations}; however, we emphasize
that---unlike our method---these baselines do not leverage insights from the literature on doubly robust
survival analysis to target treatment effects.
Moreover, following \cite{athey2016recursive}, our method directly targets heterogeneity in the conditional average treatment effect when choosing where to place splits.
In Figure \ref{fig:figure1} we plot estimated effects $\htau(X_i)$ against the true $\tau(X_i)$ on randomly
drawn test set points across two simulation designs discussed further in Section \ref{sec:simulations}.
In both cases, we find our approach to be considerably more accurate than the non-doubly robust baselines here.

Our paper is structured as follows. First, in Section~\ref{sec:csf}, we motivate and present our proposed method, causal survival forests.
In Section~\ref{sec:theory}, we study asymptotics of causal survival forests in a generalized random forest setting introduced by \cite{wager2019grf}.
In Section~\ref{sec:simulations}, we conduct simulation experiments, and find the proposed method to perform well in a wide variety of settings relative to other proposals in the literature.
In Section~\ref{sec:realdata}, our approach is illustrated via a data application to the data from AIDS Clinical Trials Group Protocol 175 (ACTG175).
A software implementation is provided as part of the package \texttt{grf} for R and C++,
available from CRAN and at \url{github.com/grf-labs/grf} \citep{GRF,CRAN}

\subsection{Related work}

As discussed above, there is an extensive recent literature on non-parametric methods for heterogeneous
treatment effect estimation \citep{athey2016recursive,wager2019grf,foster2011subgroup,friedberg2018local,hahn2017bayesian,
hill2011bayesian,kennedy2020optimal,kunzel2019metalearners,ishwaran2018estimating,luedtke2016super,nie2020quasi,oprescu2019orthogonal,
tian2014interaction,wager2018estimation}. However, right-censored survival data are frequently encountered in clinical trials
and other biomedical research studies, and challenges related to this setting have mostly not been addressed in the
existing literature. Here, we take a step towards enabling flexible heterogeneous treatment effect estimation with
survival data by adapting the forest-based method of \citet{wager2019grf} to this setting.

Random forests were originally proposed by \citet{breiman2001random}, and trace back to the work of
\citet{breiman1984classification} on CART trees. By now, the literature on general random forest methods for
survival data has already received considerable attention; and in particular
\citet{leblanc1993} and \citet{hothorn2006unbiased} studied survival tree models in the context of conditional inference trees;
\citet{hothorn2006survival} proposed using inverse probability of censoring weighting to compensate censoring in forest models;
\citet{ishwaran2008random} proposed random survival forests, which extend random forests to handle survival data via using log-rank tests at each split on an individual survival tree \citep{CIAMPI1986185,Segal1988};
\citet{zhu2012recursively} studied the impact of recursive imputation of survival forests on model fitting;
\citet{steingrimsson2016doubly} proposed doubly robust survival trees by constructing doubly robust loss functions that use more information to improve efficiency; and
\citet{steingrimsson2019} constructed censoring unbiased regression trees and forests by considering a class of censoring unbiased loss functions.
However, none of these methods were directly targeted at heterogeneous treatment effects in observational studies.

Finally, we note that the problem of heterogeneous treatment effect estimation is closely related to that of
optimal treatment regimes or policy learning \citep{Athey2019policy,Cui2021Individualized,manski2004statistical,murphy2003optimal,
qian2011performance,luedtke2016statistical,zhang2012robust, zhao2012estimating}. And, unlike in the case of heterogeneous treatment
effects, there has been more work on developing methods for optimal treatment regimes under censoring.
Adapting the outcome weighted learning framework \citep{zhao2012estimating}, \citet{zhao2015doubly} proposed two new approaches, inverse censoring weighted outcome weighted learning, and doubly robust outcome weighted learning, both of which require semi-parametric estimation of the conditional censoring probability given the patient characteristics and treatment choice. \citet{zhu2017greedy} adopted the accelerated failure time model to estimate an interpretable single-tree treatment decision rule. \citet{cui2017tree} proposed a random forest approach for right-censored outcome weighted learning, which avoids both the inverse probability of censoring weighting and restrictive modeling assumptions. However, for observational studies with censored survival outcomes, these methods may suffer from confounding and selection bias. In addition, none of the above approaches focus on estimating the heterogeneous treatment effect or on associated uncertainty quantification.

\section{Causal survival forests}
\label{sec:csf}

As discussed above, our statistical setting is specified in terms of the distribution of i.i.d. tuples
$\cb{X_i, \,  T_i, \, C_i, \, W_i} \in \xx \times \RR_+ \times \RR_+ \times \cb{0, \, 1}$, and we are interested
in measuring the effect of a binary treatment $W$ on some deterministic transformation of the survival time $T$, i.e.,
\begin{equation}
    \label{eq:rmst}
\tau(x) = \EE{y(T_i(1)) - y(T_i(0)) \cond X_i = x}.
\end{equation}
To streamline our presentation, with a slight abuse of notation, $\tau(x)$ is redefined as \eqref{eq:rmst} hereafter.
We observe censored survival times $U_i = T_i \land C_i$, and write the non-censoring
indicator $\Delta_i = 1\cb{T_i \leq C_i}$.

\begin{rema}
Note that \eqref{eq:cate} is a special case of \eqref{eq:rmst} with $y(T)=T$.
Other transformations include $y(T)=T\land h$ for the restricted mean survival time and $y(T)=1\cb{T \geq h}$ for the survival probability.
\end{rema}

Throughout, we will assume
that the transformation $y(\cdot)$ is indifferent to survival beyond some maximal time horizon
$h$. The reason we make this assumption is that, in most studies, all units are censored after some
finite amount of time (e.g., in a 10-year study, no units will be observed past the 10-year mark), and
functionals of $T_i$ that depend on behavior past this point will not be identified.
The upshot of Assumption \ref{assu:h} is that we can treat any sample tracked past the horizon $h$
as ``observed'', even if they are eventually censored.

\begin{assu}\emph{(Finite horizon)}
\label{assu:h}
The outcome transformation $y(\cdot)$ admits a maximal horizon $0 < h < \infty$, such that $y(t) = y(h)$ for all $t \geq h$.
\end{assu}

\begin{defi}\emph{(Effective non-censoring indicator)}
\label{defi:Dh}
Under Assumption \ref{assu:h}, we define the effective non-censoring indicator as $\Delta_i^h = 1\cb{(T_i \land h) \leq C_i}$; or, equivalently in terms of the observation $U_i$, we have $\Delta_i^h = \Delta_i \lor 1\cb{U_i \geq h}$.
\end{defi}

In order to identify treatment effects, we need to rely on two sets of assumptions. First, we
need to make basic causal assumptions that would enable us to estimate the causal effect of $W_i$ on
$T_i$ without censoring. The three assumptions below do so following \citet{rosenbaum1983central},
and correspond exactly to standard assumptions used to identify the conditional average treatment effect \eqref{eq:cate}
in the literature on heterogeneous treatment effect estimation \citep{kunzel2019metalearners,nie2020quasi}.

\begin{assu}\emph{(Potential outcomes)}
\label{assu:PO}
There are potential outcomes $\cb{T_i(0), \, T_i(1)}$ such that $T_i = T_i(W_i)$ almost surely.
\end{assu}

\begin{assu}\emph{(Ignorability)}
\label{assu:unconf}
Treatment assignment is as good as random conditionally on covariates, $\{ T_i(0), T_i(1)\} \perp W_i \mid X_i$.
\end{assu}

\begin{assu}\emph{(Overlap)}
\label{assu:overlap}
The propensity score $e(x) = \PP{W_i = 1 \cond X_i=x}$ is uniformly bounded away from 0 and 1,
i.e., $\eta_e \leq e(x) \leq 1 - \eta_e$ for some $0 < \eta_e \leq \frac{1}{2}$.
\end{assu}

Second, we need assumptions to guarantee that censoring due to $C_i$ does not break identification
results for treatment effects obtained via Assumptions \ref{assu:PO}--\ref{assu:overlap}. To this end, we
rely on standard assumptions from survival analysis \citep[e.g.,][]{fleming2011counting}.

\begin{assu}\emph{(Ignorable censoring)}
\label{assu:censor}
Censoring is independent of survival time conditionally on treatment and covariates, $T_i \indep C_i \mid X_i, \, W_i$.
\end{assu}

\begin{assu}\emph{(Positivity)}
\label{assu:positivity}
%The survival times $T_i$ are bounded from above by $T_{\max} \in \RR_+$ almost surely, and
$\PP{C_i < h \cond X_i, \, W_i} \leq 1 - \eta_C$ for some $0 < \eta_C \leq 1$.
\end{assu}

Assumptions~\ref{assu:censor} and \ref{assu:positivity} play a fundamental role for identification.
Qualitatively, we note that these two assumptions can be seen as direct analogues to
Assumptions~\ref{assu:unconf} and \ref{assu:overlap}, where the former control how censoring due to $C_i$
can mask information about $T_i$ whereas the latter control how treatment assignment $W_i$ can mask information
about the potential outcomes $\cb{T_i(0), \, T_i(1)}$.

Below, we start by reviewing the causal forest algorithm of \citet{wager2019grf} that can be used to estimate
the conditional average treatment effect without censoring under Assumptions~\ref{assu:PO}--\ref{assu:overlap}.
Next, we discuss two adaptations of causal forests for censored survival data: First, in Section~\ref{sec:ipcw},
we discuss a simple proposal based on inverse-probability of censoring weighting and then, in Section~\ref{sec:csf_dr},
we give our main proposal based on a doubly-robust censoring adjustment.

\subsection{Causal forests without censoring}
\label{sec:cf}

The basic causal forest algorithm is motivated by the following fact.
If we knew that the treatment effects were constant, i.e., $\tau(x) = \tau$ for all $x$, then the following
estimator $\hat{\tau}$ due to \citet{robinson1988} attains $1/\sqrt{n}$ rates of convergence, provided the three assumptions
detailed above hold and that we estimate nuisance components sufficiently fast \citep{chernozhukov2018}:
\begin{equation}
\label{eq:robinson}
\begin{split}
&\sum_{i = 1}^n \psi^{(c)}_{\hat{\tau}}(X_i, \, y(T_i), \, W_i; \, \hat{e}, \, \hat{m}) = 0, \\
&\psi^{(c)}_\tau(X_i, \, y(T_i), \, W_i; \, \hat{e}, \, \hat{m}) = \left[W_i - \hat{e}(X_i)\right]\left[y(T_i) - \hat{m}(X_i) - \tau  \left( W_i - \hat{e}(X_i) \right)\right],
\end{split}
\end{equation}
where $e(x) = \PP{W_i = 1|X_i=x}$, $m(x) = \EE{y(T_i) | X_i = x}$, and $\hat{e}(X_i)$ and $\hat{m}(X_i)$ are
estimates of these quantities derived via cross-fitting \citep{schick1986asymptotically}.
We use the superscript $(c)$ to remind ourselves that this estimator requires access to the complete
(uncensored) data.

Here, however, our goal is not to estimate a constant treatment effect $\tau$, but rather to fit
covariate-dependent treatment heterogeneity $\tau(x)$;
and we do so using
a random forest method.
As background, recall that given a target point $x$, tree-based methods seek to find training examples which are close to $x$ and use the local kernel weights to obtain a weighted averaging estimator. An essential ingredient of tree-based methods is recursive partitioning of the covariate space $\XX$, which induces the local weighting. When the splitting variables are adaptively chosen, the width of a leaf can be narrower along the directions where the causal effect is changing faster. After the tree fitting is completed, the closest points to $x$ are those that fall into the same terminal node as $x$. The observations that fall into the same node as the target point $x$ can be treated asymptotically as coming from a homogeneous group.

\cite{wager2019grf} generalizes the use of random forest-based weights for generic kernel estimation.
The most closely related precedent from the perspective of adaptive nearest neighbor estimation are quantile regression forests
\citep{meinshausen2006quantile} and bagging survival trees \citep{hothorn2004bagging}, which can be viewed as special cases
of generalized random forests. The idea of adaptive nearest neighbors also underlies theoretical analyses of random forests such as
\citet{arlot2014analysis}, \citet{biau2008consistency}, and \citet{lin2006random}.
The random forest-based weights $\alpha_i$ are derived from the fraction of trees in which an observation
appears in the same terminal node as the target point.
Specifically, given a test point $x$, the weights $\alpha_i(x)$ are the frequency with which the $i$-th training example falls in the same leaf as $x$, i.e.,
\begin{equation}
\label{eq:alpha}
\alpha_i(x) = \frac{1}{B}\sum_{b=1}^B \frac{1\cb{X_i\in \Node_b(x)}}{|\Node_b(x)|},
\end{equation}
where $\Node_b(x)$ is the terminal node that contains $x$ in the $b$-th tree, $B$ is the number of trees, and $|\cdot|$ denotes the cardinality.

The crux of the causal forest algorithm presented in
\citet{wager2019grf} is to pair the kernel-based view of forests with Robinson's estimating equation \eqref{eq:robinson}.
Causal forests seek to grow a forest such that the resulting weighting function $\alpha_i(x)$ can be used to express heterogeneity
in $\tau(\cdot)$, meaning that $\tau(\cdot)$ is roughly constant over observations given positive
weight $\alpha_i(x)$ for predicting at $x$. Then, we estimate $\tau(x)$ by solving a localized version
of \eqref{eq:robinson}:
\begin{equation}
\label{eq:grf}
\sum_{i = 1}^n \alpha_i(x) \psi^{(c)}_{\hat{\tau}(x)}(X_i, \, y(T_i), \, W_i; \, \hat{e}, \, \hat{m}) = 0.
\end{equation}
\citet{wager2019grf} discuss additional details including the choice of a splitting rule targeted for treatment heterogeneity,
while \citet{nie2020quasi} and \citet{kennedy2020optimal} further explore the idea of using Robinson's transformation to fit
treatment heterogeneity. For a general review and discussion of causal forests, see \citet{athey2019estimating}.

\subsection{Adjusting for censoring via weighting}
\label{sec:ipcw}

In the presence of censoring, the estimator \eqref{eq:grf} is no longer applicable because $T_i$ is not always
observed. Simply ignoring censoring and running causal forests with complete observations (i.e., those with $\Delta_i^h = 1$)
would lead to bias. However, under Assumptions \ref{assu:censor} and \ref{assu:positivity}, several general approaches
for making statistical estimators robust to censoring are available.

One particularly simple censoring adjustment is via inverse-probability of censoring weighting (IPCW). Let
\begin{equation}
\label{eq:SC}
S_{w}^C(s|x) = \PP{C_i \geq s |W_i=w, X_i = x}
\end{equation}
denote the conditional survival function for the censoring process\footnote{
The conditional survival function for the censoring process is defined using weak inequality
because if the censoring event occurs at the same time as the failure event, the failure event is observed.}, and note that any given survival
time is observed in the sense of Definition \ref{defi:Dh} with probability
\begin{equation}
\label{eq:PCW}
\PP{\Delta_i^h = 1 \cond X_i, \, W_i, \, T_i} = S_{W_i}^C\p{T_i \land h \cond X_i},
\end{equation}
The main idea of IPCW estimation is to only consider complete cases, but up-weight all complete observations
by $1/S_{W_i}^C\p{T_i\land h \cond X_i}$ to compensate for censoring. As discussed in \citet[Chapter 3.3]{van2003unified},
such IPCW estimators succeed in eliminating censoring bias under considerable generality.

In the context of causal forests, IPCW estimation for an outcome transformation satisfying
Assumption~\ref{assu:h} amounts to estimating $\tau(x)$ as
\begin{align}
\label{eq:ipcw-cf}
&\sum_{\cb{i : \Delta_i^h = 1}} \frac{\alpha_i(x)}{\hat{S}_{W_i}^C\p{T_i \land h\cond X_i}} \psi^{(c)}_{\hat{\tau}(x)}(X_i, \, y(T_i), \, W_i; \, \hat{e}, \, \hat{m}) = 0,
\end{align}
where $\hat{S}_w^C(s|x)$ is an estimate of \eqref{eq:SC} that's pre-computed using cross-fitting just like $\hat{e}$ and
$\hat{m}$, and $\alpha_i(x)$ is obtained as in \eqref{eq:alpha} trained on complete observations only. This algorithm forms a first adaptation of causal forests for censored data. We note that,
when using \eqref{eq:ipcw-cf}, the nuisance component $\hat{m}(x)$ also needs to be estimated in
a way that accounts for censoring; here, the IPCW approach can again be used. Concretely, we implement
IPCW causal forests by training a model on samples with observed failure times only $(\Delta_i^h=1)$, and pass $1/\hat{S}_{W_i}^C\p{T_i\land h \cond X_i}$ as a ``sample weight'' to the function
\texttt{causal\_forest} in \texttt{grf} \citep{GRF}. We refer to the \texttt{grf} package documentation for more details of how sample weights
are incorporated in all parts of the algorithm (including in splitting).

\subsection{A doubly robust correction}
\label{sec:csf_dr}

While the IPCW approach discussed above is attractive in terms of its simplicity, it has several statistical
drawbacks. First, this estimation strategy requires us to effectively throw away all observations with $\Delta_i^h = 0$,
and this may hurt us in terms of efficiency: After all, if we see that a sample got censored at time $U_i > 0$, then
at least we know they didn't die before $U_i$, and a powerful estimation procedure should be able to take
this into account. Second, in general, we need to run IPCW with an estimate of $\hat{S}_w^C(s|x)$, and
IPCW-type methods are generally not robust to estimation errors in this quantity---which will be a problem
especially if we want to use flexible methods like random survival forests for estimating $\hat{S}_w^C(s|x)$
\citep{chernozhukov2018,van2011targeted}.

For this reason, our preferred causal survival forest method does not rely on IPCW, and instead relies
on a more robust approach to making estimating equations robust to censoring, described
in  \citet[][Chapter 10.4]{tsiatis2007semiparametric}.
If the true value $\tau$ of our parameter of interest is identified by a complete
data estimating equation, $\mathbb{E}[\psi_{\tau}^{(c)}(X_i, \, y(T_i), \, W_i)] = 0$ then,
under Assumption \ref{assu:h}, $\tau$ is also identified via
the following estimator that generalizes the celebrated augmented inverse-propensity weighting estimator
of \citet{robins1994estimation}.
We have $\EE{\psi_{\tau}(X_i, \, y(U_i), \, U_i\land h, \, W_i, \, \Delta_i^h)} = 0$ with scores
\begin{equation}\label{eq:SurvExp}
\begin{split}
&\psi_\tau(X_i, \, y(U_i), \, U_i\land h, \, W_i, \, \Delta_i^h) = \frac{\Delta_i^h \,  \psi_{\tau}^{(c)}(X_i, \, y(U_i), \, W_i)}{S_{W_i}^C( U_i\land h | X_i)} \\
&\ \ \ \ \ \ \ \  + \frac{(1 - \Delta_i^h) \EE{\psi_{\tau}^{(c)}(X_i, \, y(T_i), \, W_i) | T_i\land h > U_i \land h, \, W_i, \, X_i}}{S_{W_i}^C(U_i\land h | X_i)} \\
&\ \ \ \ \ \ \ \ - \int_{0}^{U_i\land h} \frac{\lambda_{W_i}^C(s | X_i)}{S_{W_i}^C(s | X_i)}  \EE{\psi_{\tau}^{(c)}(X_i, \, y(T_i), \, W_i) | T_i\land h > s,\, W_i, \, X_i} \ ds,
\end{split}
\end{equation}
where $S_{w}^C(s|x)$ is the conditional survival function as defined in \eqref{eq:SC} and
\begin{equation}
\lambda_{w}^C(s|x) = -\frac{d}{ds} \log S_w^C(s|x)
\end{equation}
is the associated conditional hazard function.

When applied to the complete data estimating equation \eqref{eq:robinson}
used to motivate causal forests, this approach gives us scores
\begin{equation}
\label{eq:score}
\begin{split}
&\psi_\tau(X_i, \, y(U_i), \, U_i\land h, \, W_i, \, \Delta_i^h; \, \hat{e}, \, \hat{m}, \, \hat{\lambda}^C_w, \, \hat{S}^C_w, \,\hat{Q}_w ) \\
&\ \ \ = \bigg(\frac{\hat{Q}_{W_i}(U_i\land h | X_i) + \Delta_i^h [y(U_i) - \hat{Q}_{W_i}(U_i\land h | X_i)] - \hat{m}(X_i) - \tau\left( W_i - \hat{e}(X_i)\right) }{\hat{S}_{W_i}^C(U_i\land h | X_i)} \\
&\ \ \ \ \ \ - \int_{0}^{U_i \land h} \frac{\hat \lambda_{W_i}^C(s | X_i)}{\hat S_{W_i}^C(s | X_i)} [\hat{Q}_{W_i}(s | X_i) - \hat{m}(X_i) - \tau\left(W_i - \hat{e}(X_i)\right)] \ ds  \bigg)  \left(W_i - \hat{e}(X_i)\right),
\end{split}
\end{equation}
where $Q_w(s|x) = \EE{y(T_i)\cond X_i = x, \, W_i = w, \, T_i\land h > s}$ is the conditional expectation of the transformed survival time, while
$\hat{Q}_{w}(s | x)$, $\hat{S}_{w}^C(s|x)$ and $\hat{\lambda}_{w}^C(s|x)$ are cross-fit nuisance parameter estimates.
For example, $\hat{Q}_{w}(s | x)$ can be estimated by an integral of estimated survival functions and $\hat{\lambda}_{w}^C(s|x)$ can be estimated as a forward difference of $-\log(\hat{S}_{w}^C(s|x))$.

The standard way of using scores is to construct a Neyman-orthogonal robust estimator
of a relevant global parameter; see \citet{chernozhukov2018} for a general discussion and
references. In our case, the scores \eqref{eq:score} induce a robust estimator $\hat{\tau}$ for
a global constant treatment effect parameter $\tau$:
\begin{equation}\label{eq:score2}
\sum_{i = 1}^n \psi_{\hat{\tau}}(X_i, \, y(U_i), \, U_i\land h, \, W_i, \, \Delta_i^h; \, \hat{e}, \, \hat{m}, \, \hat{\lambda}^C_w,  \, \hat{S}^C_w, \, \hat{Q}_w) = 0.
\end{equation}
This estimator is Neyman-orthogonal in the sense discussed in \citet{chernozhukov2018}, and attains
a $1/\sqrt{n}$ rate of convergence for $\tau$ under 4-th root rates for the nuisance components,
provided we use cross-fitting and that Assumptions~\ref{assu:PO}--\ref{assu:positivity} hold.
Then the following result, given here for reference, illustrates the type of results one can
get from this approach. The proof follows from arguments made in
\citet{chernozhukov2018} and \citet{tsiatis2007semiparametric}; see also \citet{van2003unified} for an
extensive discussion of double robustness under right-censoring.

\begin{prop}
\label{prop:DR}
Assume that the conditional average treatment effects are constant, $\tau(x) = \tau$ for all $x \in \XX$,
and let $\tilde \tau$ be an oracle estimator for $\tau$, i.e., the solution to equation~
\eqref{eq:score2} with estimated nuisance components
$\hat{e}, \, \hat{m}, \, \hat{\lambda}^C_w/\hat{S}^C_w,  \, \hat{S}^C_w, \, \hat{Q}_w$
replaced by true values for ${e}, \, {m}, \, {\lambda}^C_w/{S}^C_w,  \, {S}^C_w, \, {Q}_w$.
Suppose
$\sup_{x\in \XX} |\hat e(x)-e(x)|=o_p(1)$, $\sup_{x\in \XX} |\hat m(x)-m(x)|=o_p(1)$, and $\sup_{x\in \XX,s\leq h}|\hat S^C_{w}(s|x)-S^C_{w}(s|x)|=o_p(1)$,
%$\sup_{x\in \XX,s\leq h} |\hat S^T_{w}(s|x)-S^T_{w}(s|x)|=o_p(1)$,
$\sup_{x\in \XX,s\leq h} |\hat Q_{w}(s|x)-Q_{w}(s|x)|=o_p(1)$,
$\sup_{x\in \XX,s\leq h} |\frac{\hat \lambda^C_{w}(s|x)}{\hat S^C_{w}(s|x)}-\frac{\lambda^C_{w}(s|x)}{ S^C_{w}(s|x)}|=o_p(1)$ for both $w=0,1$.
In addition, we assume that $\EE{|\hat e(X)-e(X)|^2}=o(b_n^2)$, $\EE{|\hat m(X)-m(X)|^2}=o(c_n^2)$,  and $\sup_{s\leq h} \EE{|\hat S^C_{w}(s|X)-S^C_{w}(s|X)|^2}=o(c_n^2)$,
%$\sup_{s\leq h} \EE{|\hat S^T_{w}(s|X)-S^T_{w}(s|X)|^2}=o(c_n^2)$,
$\sup_{s\leq h} \EE{|\hat Q_{w}(s|X)-Q_{w}(s|X)|^2}=o(c_n^2)$,
$\sup_{s\leq h} \EE{|\frac{\hat \lambda^C_{w}(s|X)}{\hat S^C_{w}(s|X)}-\frac{\lambda^C_{w}(s|X)}{ S^C_{w}(s|X)}|^2}=o(d_n^2)$ for both $w=0,1$.
Then, provided Assumptions~\ref{assu:PO}--\ref{assu:positivity} hold,
we have $\hat \tau- \tilde \tau= o_p(\max((c_n+d_n)c_n,b_n c_n,b^2_n))$.
Furthermore, the solution $\hat{\tau}$ to estimating equation~
\eqref{eq:score2} has an asymptotically normal sampling distribution
and attains a $1/\sqrt{n}$ rate of convergence, provided the nuisance components
$\hat{e}, \, \hat{m}, \, \hat{\lambda}^C_w/\hat{S}^C_w,  \, \hat{S}^C_w, \, \hat{Q}_w$ are learned using cross-fitting, are uniformly consistent and
achieve 4-th root rates of convergence in root-mean squared error.
\end{prop}

Our causal survival forests also use the doubly robust scores \eqref{eq:score}, but now in the context of a
forest-based estimator for treatment heterogeneity.
First, we estimate the nuisance components
$ \hat{e}, \, \hat{m}, \, \hat{\lambda}^C_w,  \, \hat{S}^C_w, \, \hat{Q}_w$ required to form the score \eqref{eq:score},
and then pair this estimating equation with the forest weighting scheme \eqref{eq:grf}, resulting in estimates
$\hat\tau(x)$ characterized by
\begin{equation}
\label{eq:csf}
\sum_{i = 1}^n \alpha_i(x)  \psi_{\hat{\tau}(x)}(X_i, \, y(U_i), \, U_i\land h, \, W_i, \, \Delta_i^h; \, \hat{e}, \, \hat{m}, \, \hat{\lambda}^C_w,  \, \hat{S}^C_w, \, \hat{Q}_w) = 0.
\end{equation}
In order to use this estimator, we of course need to specify how to grow the forest, so that the resulting forest weights
$\alpha_i(x)$ adequately express heterogeneity in the underlying signal $\tau(x)$.
Here, for the splitting rule, we use the $\widetilde \Delta$-criterion proposed in \citet{wager2019grf}. In particular, we generate pseudo-outcomes by the following relabeling strategy at each parent node $\Node$,
\begin{align}\label{eq:rho}
\rho_i =  \psi^i_{\hat{\tau}_\Node} \times \left[  \frac{1}{|\{j:X_j \in \Node\}|} \sum_{j: X_j\in \Node} (W_j- \hat e(X_j))^2 \left(\frac{1}{\hat{S}^C_{W_j} (U_j\land h|X_j)}-\int_0^{U_j\land h} \frac{\hat{\lambda}_{W_j}^C(s|X_j)}{\hat{S}^C_{W_j} (s|X_j)} ds\right)  \right]^{-1},
\end{align}
where $\psi^i_{\hat{\tau}_\Node}$ is a shorthand of $\psi_{\hat{\tau}_\Node}(X_i, \, y(U_i), \, U_i\land h, \, W_i, \, \Delta_i^h; \, \hat{e}, \, \hat{m}, \, \hat{\lambda}^C_w,\, \hat{S}^C_w, \, \hat{Q}_w)$, and $\hat{\tau}_\Node$ is the estimated $\tau$ in $\Node$ by \eqref{eq:score2}.
Next, the splitting criterion proceeds exactly the same as a regression tree \citep{breiman1984classification} problem by treating the pseudo-outcomes $\rho_i$'s as a continuous outcome variable. Specifically, we split the parent node into two child nodes $\Node_L$ and $\Node_R$ so as to maximize the following quantity:
\begin{align*}
\widetilde \Delta(L,R) = \frac{1}{|\{i:X_i \in \Node_L\}|} \Bigg(\sum_{i:X_i \in \Node_L} \rho_i\Bigg)^2 + \frac{1}{|\{i:X_i \in \Node_R\}|} \Bigg(\sum_{i:X_i \in \Node_R} \rho_i \Bigg)^2.
\end{align*}

\begin{rema}
\label{rema:outcomes}
Throughout this paper, we will present our method and formal results in a general setting
that covers any outcome transformation $y(\cdot)$ satisfying Assumption~\ref{assu:h}.
However, the implementation of our doubly robust method, provided as the \texttt{causal\_survival\_forest}
function in \texttt{grf}, only considers two target outcomes: The restricted mean survival time (RMST),
\begin{equation}
\tau^h(x) = \EE{T_i(1) \land h \cond X_i = x}-\EE{T_i(0) \land h \cond X_i = x},
\end{equation}
which arises from our framework using $y(T) = T \land h$, and the survival probability,
\begin{equation}
\pi^h(x) = \PP{T_i(1) \geq h \cond X_i = x}-\PP{T_i(0) \geq h \cond X_i = x},
\end{equation}
which arises from using $y(T) = 1\cb{T \geq h}$. Methodological extensions to other outcome
transformations is straightforward, but requires modifying estimators for the nuisance components
$\hat{Q}_w(s|x)$ and $\hat m(x)$. In our implementation for estimating $\tau^h(x)$ and $\pi^h(x)$, $\hat{Q}_w(s|x)$ and $\hat m(x)$ are derived from the estimated conditional survival function of the survival time.
\end{rema}

\begin{rema}
\label{rema:splitting}
Our splitting rule based on pseudo-outcomes as in \eqref{eq:rho} is a direct
application of the abstract GRF-algorithm of \citet{wager2019grf}. This algorithmic technique
is motivated by classical influence-function based approximations
for semiparametric inference \citep{andrews1993tests,zeileis2005unified}.
An alternative approach to splitting that could
also be used with our doubly robust estimating equation is to maximize a test statistic
for treatment heterogeneity as in, e.g., \citet{zeileis2008model} or \citet{yang2021causal}.
\citet{wager2019grf} have a further discussion of both statistical and computational
properties of pseudo-outcome-based splitting.
\end{rema}

\section{Asymptotics and inference}
\label{sec:theory}

A main draw of forest based methods is that they have repeatedly proven themselves to be both robust and
flexible in practice. However, to further our statistical understanding of the method, it is helpful to study
the behavior of the method under a more restricted asymptotic setting where sharp formal characterizations
are available. To this end, we now study asymptotics of causal survival forests in a setting that builds on the
one used by \citet{wager2018estimation} to study inference with forests in problems without censoring.
Throughout this section, we assume that the covariates $X \in \XX=[0,1]^p$ are distributed according to a density that is
bounded away from zero and infinity. The following assumption guarantees the smoothness of $\EE{\psi_{\tau(x)}|X=x}$.

\begin{assu}\label{assumption:Lips}
(Lipschitz continuity)
The treatment effect function $\tau(x)$ is Lipschitz continuous in terms of $x$. In addition, the nuisance components $e(x)$, $m(x)$ are Lipschitz continuous in terms of $x$, and $Q_w(s|x)$, $\lambda_w^C(s|x)$, $S_w^C(s|x)$ are Lipschitz continuous in terms of $x$ for both $w\in \{0,1\}$ and all $s\leq h$.
\end{assu}

In addition, our trees are symmetric, i.e., their output is invariant to permuting the indices of training samples.
Our algorithm also guarantees honesty \citep{wager2018estimation}, and the following two conditions.

%\begin{assu}
\emph{Random split tree:}
At each internal node, the probability of splitting at the $j$-th dimension is greater than $\varsigma$, where $0<\varsigma<1$ for $j=1,\ldots,p$.
%\end{assu}

%\begin{assu}
\emph{Subsampling:}
Each child node
contains at least a fraction $\nu$ of the data points in its parent node for some $0<\nu<0.5$, and trees are grown on subsamples of size $\ell$ scaling as
\begin{equation}
\label{eq:subsamp}
\ell=n^\gamma, \ \ \ \ \kappa<\gamma<1, \ \ \ \ \kappa \equiv 1-\big[1+\varsigma^{-1} \big( \log(\nu^{-1})\big)/ \big(\log(1-\nu)^{-1}\big) \big]^{-1}.
\end{equation}
We note that, in Theorem \ref{theo:pointwise}, we will obtain a rate of convergence of $\widetilde{O}(\sqrt{\ell / n})$ for $\htau(x)$, where $\widetilde{O}(\cdot)$ denotes the term neglecting the logarithmic factors; thus \eqref{eq:subsamp}
captures the asymptotic behavior we can get; see \citet{wager2018estimation} for an
in-depth discussion of this assumption.
%\end{assu}

Moreover, we need  Assumption~\ref{assumption:plug-in} to couple $\hat \tau(x)$ and  $\tilde \tau(x)$, where $\tilde \tau(x)$ is an oracle estimator, with nuisance components
$ {e}, \, {m}, \, {\lambda}^C_w,  \, {S}^C_w, \, {Q}_w$ being the underlying truth instead of their empirical analogues in equation~\eqref{eq:score}.

\begin{assu}
\label{assumption:plug-in}
Consistency of the non-parametric plug-in estimators: we have the following convergences in probability,
\begin{align*}
 \sup_{x\in \XX} |\hat e(x)-e(x)| \rightarrow 0,~ \sup_{x\in \XX} |\hat m(x)-m(x)|\rightarrow 0,
 \end{align*}
 and
 \begin{align*}
 \sup_{x\in \XX,s\leq h}|\hat S^C_{w}(s|x)-S^C_{w}(s|x)|\rightarrow 0,
%\sup_{x\in \XX,s\leq h} |\hat S^T_{w}(s|x)-S^T_{w}(s|x)|\rightarrow 0,\\
\sup_{x\in \XX,s\leq h} |\hat Q_{w}(s|x)-Q_{w}(s|x)|\rightarrow 0,
\sup_{x\in \XX,s\leq h} |\frac{\hat \lambda^C_{w}(s|x)}{\hat S^C_{w}(s|x)}-\frac{\lambda^C_{w}(s|x)}{ S^C_{w}(s|x)}|\rightarrow 0,
\end{align*}
for both $w=0,1$. Furthermore, suppose that \begin{align*}
\EE{\sup_{x\in \XX}|\hat e(x)-e(x)|^2}=o(b_n^2),~ \EE{\sup_{x\in \XX}|\hat m(x)-m(x)|^2}=o(c_n^2),
\end{align*}
and
\begin{align}
\sup_{s\leq h} \EE{\sup_{x\in\XX}|\hat S^C_{w}(s|x)-S^C_{w}(s|x)|^2}=o(c_n^2),\label{eq:rate1}\\
%$\sup_{s\leq h} \EE{\sup_{x\in\XX}|\hat S^T_{w}(s|x)-S^T_{w}(s|x)|^2}=o(c_n^2)$,
\sup_{s\leq h} \EE{\sup_{x\in\XX}|\hat Q_{w}(s|x)-Q_{w}(s|x)|^2}=o(c_n^2),\label{eq:rate2}\\
\sup_{s\leq h} \EE{\sup_{x\in\XX}|\frac{\hat \lambda^C_{w}(s|x)}{\hat S^C_{w}(s|x)}-\frac{\lambda^C_{w}(s|x)}{ S^C_{w}(s|x)}|^2}=o(d_n^2)\label{eq:rate3}
\end{align}
for both $w=0,1$.
\end{assu}

Equations~\eqref{eq:rate1}-\eqref{eq:rate3} imply the corresponding rate assumptions of Proposition~\ref{prop:DR}.
Assumption~\ref{assumption:plug-in} is quite general and can be achieved by many existing models. \cite{biau2012analysis,wager2015adaptive} show that for the random forest models, $b_n^2$ can be faster than $n^{-2/(p+2)}$ as long as the intrinsic signal dimension is less than $0.54 p$. As shown in
\cite{cui2022consistency}, $c_n^2=n^{-1/(p+2)}$ is achievable for survival forest models. Non-parametric kernel smoothing methods such as \cite{10.1093/biomet/asy064} provide estimation with $d_n^2= n^{-1/2+\kappa}$, where $\kappa<1/2$ depending on the dimension $p$.

The following lemma provides an intermediate result for our main theorem, which bounds the difference between $\hat \tau(x)$ and $\tilde \tau(x)$.

\begin{lemm}
We assume Assumptions~\ref{assu:positivity} and \ref{assumption:plug-in} hold. Then for any $x\in \XX$, we have that  $\hat \tau(x)- \tilde \tau(x)= o_p(\max((c_n+d_n)c_n,b_n c_n,b^2_n))$.
\label{lemma:couple}
\end{lemm}

The proof of Lemma~\ref{lemma:couple} is deferred to the Appendix. The technical results in \cite{wager2018estimation,wager2019grf} paired with Lemma~\ref{lemma:couple} lead to the following asymptotic Gaussianity result.

\begin{theo}
\label{theo:pointwise}
Assume Assumptions~\ref{assu:PO}--\ref{assumption:plug-in} hold, and that $\ell$ scales as in \eqref{eq:subsamp}.
If $o_p(\max((c_n+d_n)c_n,b_n c_n,b^2_n))$ converges to zero with a faster rate than $polylog(n/\ell)^{-1/2}(\ell/n)^{1/2}$, where $polylog(n/\ell)$ is a function that is bounded away from 0 and increases at most polynomially with the log-inverse sampling ratio $\log (n/\ell)$.
Then there exists a sequence $\sigma_n(x)$ such that for any $x \in \XX$,
\begin{equation}
\label{eq:theo_main}
[\hat \tau(x) -\tau(x)]/\sigma_n(x) \rightarrow N(0,1),
\end{equation}
where $\sigma^2_n(x) = polylog(n/\ell)^{-1}\ell/n$.
\label{theorem:main}
\end{theo}

The proof of Theorem~\ref{theorem:main} is deferred to the Appendix. This asymptotic Gaussianity result yields valid asymptotic confidence intervals for the true treatment effect $\tau(x)$.

\subsection{Pointwise confidence intervals}
\label{sec:pointwise_ci}

One important consequence of Theorem \ref{theorem:main} is that, in settings where its assumptions hold,
we can use \eqref{eq:theo_main} to build pointwise confidence intervals for $\tau(x)$:
All we need for confidence intervals is a consistent estimator for the asymptotic variance of $\htau(x)$.
We proceed using the ``bootstrap of little bags'' algorithm \citep*{wager2019grf,sexton2009rf}. The main idea of
the bootstrap of little bags is to introduce some cluster structure into the random samples used by the random
forest to grow each tree, and then use within vs between-cluster correlations in the behaviors of individual
trees to reason about what we would have gotten by running (computationally infeasible) bootstrap  uncertainty
quantification on the whole forest.

Now, following the line of argument in Section 4 from \citet{wager2019grf}, under the conditions of
Theorem~\ref{theorem:main} our estimator $\htau(x)$ is asymptotically equivalent to
\begin{equation}
\tilde \tau^*(x) = \tau(x) + \sum_{i=1}^n \alpha_i(x) \rho_i^*(x),
\end{equation}
where $\rho_i^*(x)$ is the influence function of the $i$-th observation with respect to the true parameter value $\tau(x)$,
i.e., $\rho_i^*(x) = \psi^i_{\tau(x)}V(x)^{-1}$ with
\begin{equation}
\label{eq:V}
V(x) = \EE{ \p{W- e(X)}^2 \left(\frac{1}{S^C_{W} (U\land h|X)}-\int_0^{U\land h} \frac{\lambda_{W}^C(s|X)}{S^C_{W} (s|X)} ds \right) \bigg| X=x}.
\end{equation}
It thus follows that, to build confidence intervals, it suffices to estimate
\begin{equation}
\sigma_n^2(x) = \Var{\tilde \tau^*(x)} = \Var{\sum_{i=1}^n \alpha_i(x) \psi^i_{\tau(x)}} V(x)^{-2},
\end{equation}
where we note $\sum_{i=1}^n \alpha_i(x) \psi^i_{\tau(x)}$ is formally equivalent to the prediction made by a
regression forest with ``outcome'' $\psi^i_{\tau(x)}$.
To this end, we set
\begin{equation}
\hat \sigma_n^2 =   \hat H_n(x) \hat V_n(x)^{-2},
\end{equation}
where \smash{$\hat H_n(x)$} is estimated via the bootstrap of little bags device described above,
and \smash{$\hat V_n(x)$} is a sample-version of \eqref{eq:V} with forest-weights $\alpha_i(x)$.
We refer to Section 4 of \citet{wager2019grf} for further details and discussion of this approach.

 \subsection{Inference on linear projections of the CATE}
%\subsection{Interpreting CATE estimates}
\label{sec:BLP}

The pointwise confidence intervals discussed above provide a valuable method for assessing the stability of causal survival forests, and also can be helpful in getting a handle on the behavior of $\tau(x)$ in large samples. However, using these estimates in practice can sometimes lead to difficulties. First, the problem of pointwise non-parametric inference of $\tau(x)$ is fundamentally
a very difficult problem as the function $\tau(\cdot)$ may take on complex shapes \citep{chernozhukov2018,imai2019experimental}, which means that intervals for $\tau(x)$ will
in general be quite wide. Second, the result \eqref{eq:theo_main} relies on the forest being tuned
for ``undersmoothing'', i.e., that errors of the forest are dominated by variance and bias is negligible.
In practice, it can be difficult to detect cases where undersmoothing does not hold (and confidence intervals are centered
on potentially biased point-estimates); see Appendix C3 of \citet{wager2019grf} for further discussion.

In order to get around these difficulties, \citet{semenova2017estimation} advocate focusing
inference on low-dimensional summaries of $\tau(\cdot)$, including projections \citep{beran1977,white1980,white1982,buja2019models};
see also \citet{van2006statistical} and \citet{NEUGEBAUER2007}.
Given a set of covariates $A_i$, the best linear projection of the CATE function $\tau(\cdot)$ is
\begin{equation}
\label{eq:BLP}
\cb{\beta^*_0, \, \beta^*} = \argmin_{\beta_0,\beta} \EE{\p{\tau(X_i) - \beta_0 - A_i \beta}^2}.
\end{equation}
Typically, the $A_i$ will be chosen as a subset (or transformations) of the $X_i$.
As argued in \citet{semenova2017estimation}, such linear projections can be used to meaningfully interpret and summarize
treatment heterogeneity, but remain simple enough that we can still provide robust inference for them using
well-understood techniques from the literature on semiparametrics.
This means a researcher is able to pre-specify a hypothesis of how they believe the conditional mean of $\tau(X)$
varies with for example age and gender, and achieve valid inference for this association measure, even though point
estimates of $\tau(X)$ may be quite uncertain and obtained non-parametrically.

To implement this approach, we first construct relevant doubly robust scores, based on the augmented inverse propensity weighting \citep{robins1994estimation}
\begin{equation}
\label{eq:DR}
\hGamma_i = \hat \tau(X_i) + \frac{\psi_{\hat \tau(X_i)}(X_i, \, y(U_i), \, U_i\land h, \, W_i, \, \Delta_i^h)}{\hat e(X_i)(1-\hat e(X_i))},
\end{equation}
where $\psi_\tau(\cdot)$ is as in \eqref{eq:score} and $\hat \tau(X_i)$ are CATE estimates provided by the causal
survival forest. We then estimate the best linear projection (BLP) parameters \eqref{eq:BLP} by running a linear
regression of \smash{$\hGamma_i$} on the features of interest $A_i$; i.e., the regression $\hGamma_i \sim A_i$.
Confidence intervals can be derived via any misspecification- and heteroskedasticity-robust approach to
inference with ordinary least squares; in our implementation, we use the sandwich variance estimator with $HC_3$ covariance weights following \citet{mackinnon1985some}.

\citet{semenova2017estimation} show that this approach yields a $\sqrt{n}$-rate central limit theorem for
$\beta$ and $\beta_0$ under conditions analogous to those of Proposition \ref{prop:DR} on the nuisance
components. In the case of our forest-based approach, the rates of convergence for nuisance components
may be slower than those required by the result of \citet{semenova2017estimation} so we cannot formally
verify that a $\sqrt{n}$-rate central limit theorem holds for us; however, as shown in our simulation experiments
(see Figure~\ref{blpsim}), the estimator of \citet{semenova2017estimation} has a very nearly normal sampling
distribution.

\begin{rema}
It would also be possible to consider the pseudo-outcomes from \eqref{eq:DR} as left-hand-side variables for non-parametric regression, thus
providing an alternative starting point for non-parametric estimation of CATE with survival outcomes. This class of approaches has received
considerable attention in recent years; see, e.g., \citet{fan2022estimation}, \citet{kennedy2020optimal} and \citet{zimmert2019nonparametric}.
Meanwhile, another alternative to best linear projections is to apply assumption-lean inference recently proposed by \citet{vansteelandt2020assumption}
who use non-parametric projection to assess associations of the CATE with one variable at a time. \citet{vansteelandt2020assumption} argue
that this approach provides robust summaries of the CATE when $\tau(\cdot)$ is non-linear in terms of the summarizing variables $A_i$.
\end{rema}

\section{Simulation study}
\label{sec:simulations}

We now turn to a simulation study to assess the empirical performance of causal survival forests. As discussed in
Remark \ref{rema:outcomes}, given a target horizon $h$, we focus on estimating the effect of the treatment on both the
restricted mean survival time at $h$ and on survival probability at $h$. The choice of horizon $h$ is given in each
simulation specification.

In our experiments, we compare the following methods. {\bf (1)} An adaptation of the virtual twin {\bf (VT)} method of \citet{foster2011subgroup} to survival data, using random survival forests as implemented in the package \texttt{randomForestSRC} \citep{ishwaran2019manual}: We train two random survival forests, the first using the observations in the control group $\{(X, \, U, \, \Delta)\}_{W=0}$ to estimate $\mu_0(x) = \EE{y(T_i) \cond X_i = x, \, W_i = 0}$, and the second using the observations in the treated group to estimate $\mu_1(x) = \EE{y(T_i) \cond X_i = x, \, W_i = 1}$, where the conditional expectation estimation is constructed from the estimated conditional survival function, and then report $\htau(x) = \hmu_1(x) - \hmu_0(x)$. {\bf (2)} An instantiation of the $S$-learner strategy discussed in \citet{kunzel2019metalearners}
using random survival forests {\bf (SRC1)} as implemented in the package \texttt{randomForestSRC}:
We train a single random survival forest with features $(X, \, W)$ to estimate
$\mu(x, \, w) = \EE{y(T_i) \cond X_i = x, \, W_i = w}$, where the conditional expectation estimation is constructed from the estimated conditional survival function, and then report $\htau(x) = \hmu(x, \, 1) - \hmu(x, \, 0)$.
{\bf (3)} Enriched random survival forests {\bf (SRC2)} derived as above, except we train the forest with additional interaction
features, i.e., $(X, \, W, \, XW)$ as considered in \cite{ishwaran2018estimating}.
{\bf (4)} Inverse-propensity of censoring weighted {\bf (IPCW)} causal forests, as described in Section~\ref{sec:ipcw}.
{\bf (5)} Our proposed causal survival forests {\bf (CSF)}, as described in Section~\ref{sec:csf_dr}. Both of our methods,
IPCW and CSF, are run using publicly available implementations in the \texttt{R} package \texttt{grf} version 2.1, available from CRAN and at \url{github.com/grf-labs/grf} \citep{GRF,CRAN}.

We run the above methods on the following simulation designs. In each, we generated independent covariates from a uniform distribution on $[0,1]^p$ with $p = 15$. We consider two estimands, the restricted mean survival time (RMST) and survival probability. For the first estimand, the truncation time is listed in the simulation settings; for the second estimand, we consider surviving past the 90-th percentile of $U$. The Supplementary Material shows results when the covariates have a non-diagonal covariance matrix $V$ with entries $V_{ij} = 0.5^{|i - j|}$.

\vspace{0.2\baselineskip}
\noindent
{\it Setting 1.}
We generate $T$ from an accelerated failure time model, and $C$ from a Cox model,
\begin{align*}
&\log({T}) = -1.85 - 0.8I(X_{(1)}<0.5)+0.7X_{(2)}^{1/2}+0.2X_{(3)}  + (0.7-0.4I(X_{(1)}<0.5)-0.4X_{(2)}^{1/2})W + \epsilon, \\
&\lambda_C(t \mid W,X) = \lambda_0(t)\exp[-1.75-0.5X_{(2)}^{1/2}+0.2X_{(3)} +(1.15+0.5I(X_{(1)}<0.5)-0.3X_{(2)}^{1/2})W],\nonumber
\end{align*}
where the baseline hazard function $\lambda_0(t) = 2t$, and $\epsilon$ follows a standard normal distribution.  The follow-up time is $h=1.5$, and propensity score is $e(x)= (1+\beta(x_{(1)};2,4))/4$, where $\beta(\cdot;a,b)$ is the density function of a Beta distribution with shape parameters $a$ and $b$.

\vspace{0.2\baselineskip}
\noindent
{\it Setting 2.}
We generate $T$ from a Cox model with a non-linear structure, and $C$ from a uniform distribution on (0,3),
\begin{align*}
\lambda_{T}(t \mid W,X) =& \lambda_0(t)\exp[ X_{(1)} +(-0.5 + X_{(2)})W], \nonumber
\end{align*}
where the baseline hazard function is $\lambda_0(t) = 1/2t^{-1/2}$, and $\epsilon$ follow a standard normal distribution.
The maximum follow-up time is $h=2$, and the propensity score is $e(x)= (1+\beta(x_{(2)};2,4))/4$.

\vspace{0.2\baselineskip}
\noindent
{\it Setting 3.}
We generate $T$ from a Poisson distribution with mean $X_{(2)}^2+X_{(3)}+6+2(X_{(1)}^{1/2}-0.3)W$, and $C$ from a Poisson distribution with mean $12 +\log(1+\exp(X_{(3)}))$. The maximum follow-up time is $h=15$, and the propensity score is $e(x)= (1+\beta(x_{(1)};2,4))/4$.

\vspace{0.2\baselineskip}
\noindent
{\it Setting 4.}
We generate $T$ from a Poisson distribution with mean $X_{(2)}+X_{(3)}+\max(0,X_{(1)}-0.3)W$, and $C$ from a Poisson distribution with mean $1+\log(1+\exp(X_{(3)}))$. The maximum follow-up time is $h=3$, and propensity score is $e(x)= [(1+\exp(-x_{(1)}))(1+\exp(-x_{(2)}))]^{-1}$. Note that for subjects with $X_{(1)}<0.3$, treatment does not affect survival time, and thus the classification error rate is evaluated on the subgroup of $X_{(1)}\geq0.3$.

Default tuning parameters were used for different forest-based methods.
For the proposed causal survival forests, and for causal forests with inverse-probability of censoring weights, the default values are given in the \texttt{grf} package documentation: the number of trees is 2000, the minimum node size is 5, the subsampling fraction is 0.5, and the number of split variables $\min(p, \lceil \sqrt{p} \rceil + 20)$, where $\lceil x \rceil$ denotes the least integer greater than or equal to $x$. For estimating the nuisance components, we also use default values: 500 trees, and a minimum node size of 15 for the survival forests, and the remaining parameters, the subsampling fraction and the number of split variables are the same as that of causal survival forests.
For random survival forests \citep{ishwaran2019manual}, the minimal number of observations in each terminal node was chosen
as the default, i.e., 15;
The number of variables available for splitting at each tree node was chosen as the default,
i.e., $\lceil p^{1/2} \rceil$.
The total number of trees was set to 500.

\begin{rema}
The default tuning parameters in \texttt{grf} were chosen based on empirical performance in a series of experiments (performed independently from the research reported in this paper). This strategy reflects standard practice in applications. Another approach (not taken here) would have been to choose tuning parameters based on $n$ and $p$ using an algorithm guaranteed to satisfy asymptotic scaling conditions assumed in Theorem \ref{theorem:main}.
\end{rema}

\subsection{Results}

Table \ref{tab:MSE} reports simulation results in terms of test set mean-squared error for $\tau(X_i)$. Table \ref{tab:classif} considers classification accuracy, i.e., $1 - \frac{1}{n_{\text{test}}} \sum_{i=1}^{n_{\text{test}}} 1\cb{\sign(\htau(X_i)) = \sign(\tau(X_i))}$. These two tables report results with training set size $n = 2000$, while Figures~\ref{simCLF_RMST}-\ref{simMSE_SP} in Section~\ref{sec:simfigures} in the Appendix show results across several values of $n$.

In almost all scenarios, the proposed causal survival forest achieves the best performance among all competing methods and, overall, the proposed causal survival forest is superior to ordinary random survival forests which do not target the causal parameter directly. The exception is for setting 4 where, when targeting RMST, IPCW performs best, and when targeting the survival probability, SRC1 performs best. IPCW generally outperforms the other baselines (except for setting 4 where SRC1 is best). Two aspects of our algorithm that may help explain this finding is that, first, both IPCW and CSF are designed to target treatment effects specifically, and so are able to focus on covariates interacting with $W$ rather than on the covariates which only appear in the main effect. Second, all of these simulation processes involve non-trivial right-censoring or confounding mechanism and, as emphasized throughout the paper, CSF was designed to robustly adjust for such censoring and confounding.

\begin{table}[t]
\begin{center}
    \subcaption*{Panel A: RMST}
    \begin{tabular}{llrrrrr}
    Setting & Metric & VT & SRC1 & SRC2 & IPCW & CSF \\
      \multirow{2}{*}{1} & MSE & 0.82 & 0.46 & 0.71 & 0.60 & 0.25 \\
       & Excess MSE & 3.97 & 2.07 & 3.50 & 2.63 & 1.02 \\
      \multirow{2}{*}{2} & MSE & 2.96 & 1.89 & 2.16 & 1.75 & 1.16 \\
       & Excess MSE & 3.11 & 1.93 & 2.23 & 1.67 & 1.02 \\
      \multirow{2}{*}{3} & MSE & 49.19 & 33.31 & 42.21 & 16.88 & 13.69 \\
      & Excess MSE & 4.28 & 2.79 & 3.66 & 1.32 & 1.02 \\
      \multirow{2}{*}{4} & MSE & 5.86 & 3.79 & 4.58 & 2.79 & 3.04 \\
       & Excess MSE & 2.52 & 1.60 & 1.94 & 1.10 & 1.23 \\
    \end{tabular}

    \bigskip
    \subcaption*{Panel B: Survival probability}

    \begin{tabular}{llrrrrr}
    Setting & Metric & VT & SRC1 & SRC2 & IPCW & CSF \\
    \multirow{2}{*}{1} & MSE & 0.45 & 0.26 & 0.41 & 0.19 & 0.14 \\
       & Excess MSE & 4.16 & 2.23 & 3.92 & 1.42 & 1.07 \\
      \multirow{2}{*}{2} & MSE & 0.74 & 0.46 & 0.54 & 0.41 & 0.37 \\
      & Excess MSE & 2.41 & 1.46 & 1.72 & 1.21 & 1.06 \\
      \multirow{2}{*}{3} & MSE & 0.38 & 0.25 & 0.32 & 0.16 & 0.15 \\
      & Excess MSE & 2.86 & 1.78 & 2.42 & 1.11 & 1.05 \\
      \multirow{2}{*}{4} & MSE & 0.46 & 0.25 & 0.38 & 0.28 & 0.27 \\
      & Excess MSE & 2.06 & 1.11 & 1.70 & 1.21 & 1.21 \\
    \end{tabular}
    \end{center}

\caption{Mean squared error (MSE) multiplied by 100 and excess MSE for four scenarios, defined as
$\frac{1}{B} \sum_{b=1}^{B} \text{MSE}_b(\text{method}) / \min\{\text{MSE}_b(m) : m \in \{\text{all methods}\}\}$, respectively, for two estimands, RMST and survival probability. The simulation settings and methods under consideration are as described in Section \ref{sec:simulations}. All forests were trained on $n = 2000$ samples, and an independent test set with size $n_{\text{test}} = 2000$ was used to evaluate error. Each simulation was repeated 250 times.}
\label{tab:MSE}
\end{table}

\begin{table}[t]
\begin{center}
    \subcaption*{Panel A: RMST}
    \begin{tabular}{cccccc}
     Setting & VT & SRC1 & SRC2 & IPCW & CSF \\
    1 & 0.27 & 0.23 & 0.27 & 0.23 & 0.22 \\
    2 & 0.25 & 0.21 & 0.23 & 0.26 & 0.15 \\
    3 & 0.18 & 0.15 & 0.19 & 0.08 & 0.08 \\
    4 & 0.14 & 0.10 & 0.11 & 0.01 & 0.00 \\
    \end{tabular}

    \bigskip
    \subcaption*{Panel B: Survival probability}
    \begin{tabular}{cccccc}
     Setting & VT & SRC1 & SRC2 & IPCW & CSF \\
    1 & 0.28 & 0.25 & 0.28 & 0.22 & 0.22 \\
    2 & 0.25 & 0.22 & 0.24 & 0.25 & 0.22 \\
    3 & 0.19 & 0.16 & 0.17 & 0.09 & 0.09 \\
    4 & 0.16 & 0.11 & 0.11 & 0.05 & 0.03 \\
    \end{tabular}
    \end{center}

\caption{Classification error for two estimands, RMST and survival probability, for four scenarios, defined as $1 - \frac{1}{n_{\text{test}}} \sum_{i=1}^{n_{\text{test}}} 1\cb{\sign(\htau(X_i)) = \sign(\tau(X_i))}$ (for the last setting the classification error rate is evaluated on the subgroup of $X_{(1)}\geq0.3$). The simulation settings and methods under consideration are as described in Section \ref{sec:simulations}. All forests were trained on $n = 2000$ samples, and an independent test set with size $n_{\text{test}} = 2000$ was used to evaluate classification error. Each simulation was repeated 250 times.}
\label{tab:classif}
\end{table}

Next, we consider the accuracy of the pointwise confidence intervals for causal survival forests proposed
in Section~\ref{sec:pointwise_ci}. To do so, we evaluated the coverage of the proposed  95\%  confidence intervals at four deterministic points, namely $x_1=(0.2,\ldots, 0.2)^T$, $x_2=(0.4,\ldots, 0.4)^T$, $x_3=(0.6,\ldots, 0.6)^T$ and $x_4=(0.8,\ldots, 0.8)^T$. The true treatment effect was estimated by the Monte Carlo method with sample size 100000. The results, summarized in Table~\ref{tab:CI}, are mostly promising, and suggests that in most of these examples the bias-variance trade off of causal survival forests puts us in a regime where discussions from Section~\ref{sec:pointwise_ci} apply. However, at other points we observe poor coverage, especially at $x_1$ and $x_4$, which are near the corners of the feature space.

\begin{table}[t]
\begin{center}
    \subcaption*{Panel A: RMST}
    \begin{tabular}{rrrrrrrrr}
    & \multicolumn{4}{c}{Coverage} & \multicolumn{4}{c}{Length} \\
    Setting &   $x_1$ &   $x_2$  &  $x_3$  & $x_4$ &   $x_1$ &   $x_2$  &  $x_3$  & $x_4$ \\
    1 & 0.84 & 0.69 & 0.88 & 0.95 & 0.09 & 0.08 & 0.15 & 0.18 \\
    2 & 0.75 & 0.93 & 0.92 & 0.82 & 0.32 & 0.25 & 0.21 & 0.23 \\
    3 & 0.82 & 0.96 & 0.89 & 0.72 & 0.96 & 0.81 & 0.82 & 0.95 \\
    4 & 0.37 & 0.68 & 0.96 & 0.88 & 0.36 & 0.31 & 0.32 & 0.39 \\
    \end{tabular}

    \bigskip
    \subcaption*{Panel B: Survival probability}
    \begin{tabular}{rrrrrrrrr}
    & \multicolumn{4}{c}{Coverage} & \multicolumn{4}{c}{Length} \\
    Setting &   $x_1$ &   $x_2$  &  $x_3$  & $x_4$ &   $x_1$ &   $x_2$  &  $x_3$  & $x_4$ \\
    1 & 0.83 & 0.76 & 0.91 & 0.93 & 0.07 & 0.06 & 0.14 & 0.16 \\
    2 & 0.71 & 0.92 & 0.93 & 0.86 & 0.18 & 0.14 & 0.12 & 0.13 \\
    3 & 0.84 & 0.95 & 0.87 & 0.62 & 0.10 & 0.09 & 0.10 & 0.12 \\
    4 & 0.67 & 0.84 & 0.92 & 0.43 & 0.10 & 0.10 & 0.11 & 0.15 \\
    \end{tabular}
    \end{center}

\caption{Coverage (\%) of the proposed 95\% confidence intervals and average length at four deterministic points described in Section \ref{sec:simulations}, with a training set size $n = 2000$. $B=10000$ trees are used to fit confidence intervals. The numbers are aggregated over 1000 simulation replications.}
\label{tab:CI}
\end{table}

\begin{figure}[t]
\centering
\includegraphics[scale=0.49]{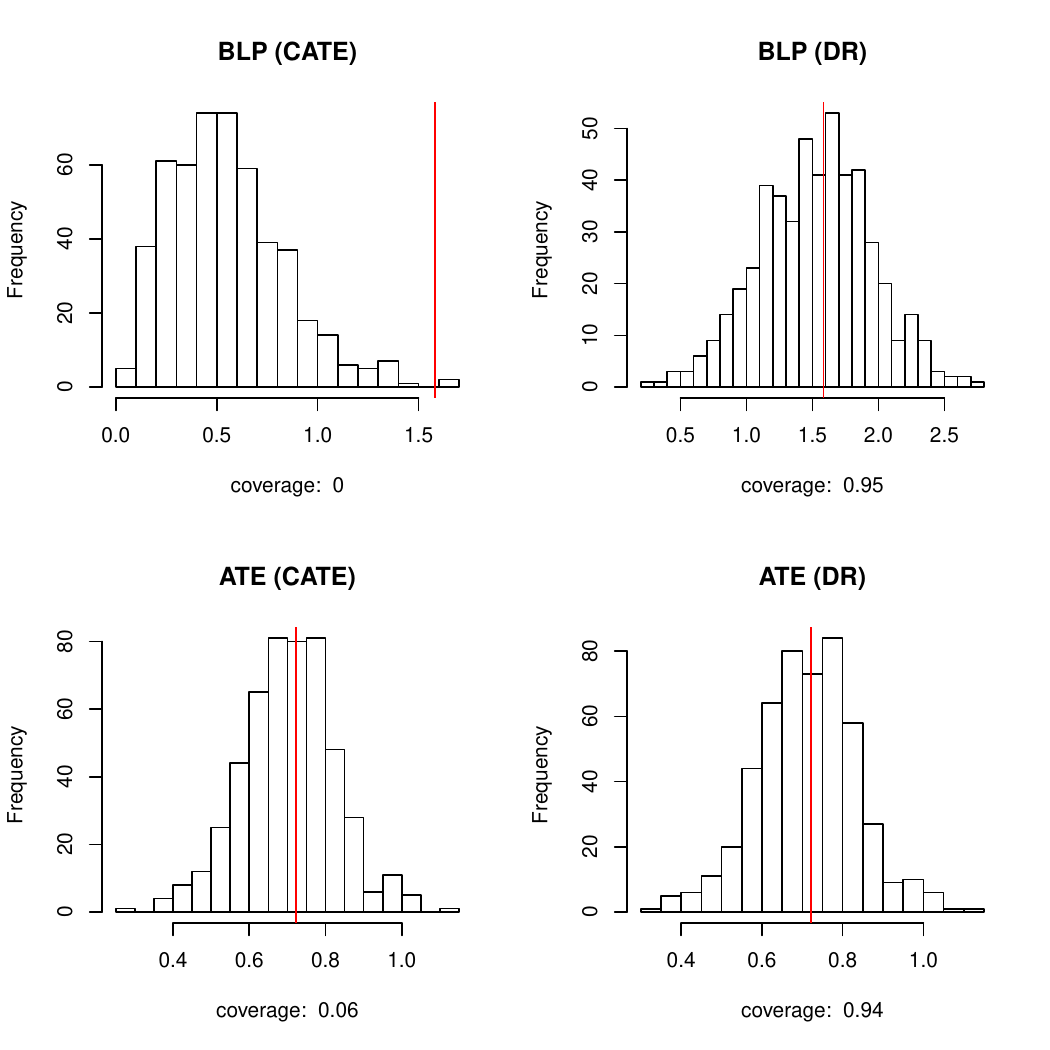}
\caption{First row: histogram of the estimated best linear projection (BLP) coefficient $\hbeta_1$ in the regression $\hat \tau(x) = \hbeta_0 + \hbeta_1 X_{(1)} + \hbeta_2 X_{(2)}$, without (left) and with (right) a robustness correction to $\hat \tau(x)$, the estimated CATEs with CSF. Second row: histogram of the estimated coefficient $\beta_0$ (ATE) in the regression $\hat \tau(x) = \hat \beta_0$. The solid red line indicates the true population coefficient. Coverage is based on 95 \% confidence intervals computed as $\hat \beta \pm z_{0.975} se(\hat \beta)$, where $z$ is standard normal quantiles and $se(\hat \beta)$ is derived via the $HC_3$ variance estimate \citep{mackinnon1985some}. Data is generated according to {\it Setting 3} with $n = 2000$ training samples and $p = 15$ covariates. The number of repetitions is 500. \label{blpsim}}
\end{figure}

Finally, we investigate the performance of the best linear projection estimator
discussed in Section \ref{sec:BLP} which provides summaries of the CATE that are amenable to more robust inference.
In Figure~\ref{blpsim}, given various choices of projection variables $A_i$ in \eqref{eq:BLP},
we consider both the doubly robust method of \citet{semenova2017estimation} (DR) and a naive baseline that simply regresses the fitted $\htau(X_i)$
estimates against $A_i$ without a doubly robust correction (CATE). We find that the proposed BLP method
performs well; in contrast, the naive direct regression can be biased (top row), and badly underestimates
the sampling variability of these estimators, thus resulting in poor coverage (both rows).

\section{HIV data analysis}
\label{sec:realdata}

We demonstrate the proposed method by an application to the data from AIDS Clinical
Trials Group Protocol 175 (ACTG175) \citep{hammer1996trial}.
The original dataset consists of 2139 HIV-infected subjects. The enrolled subjects were randomized to four treatment groups: zidovudine (ZDV) monotherapy, ZDV+didanosine (ddI), ZDV+zalcitabine, and ddI monotherapy.
We focus on the subset of patients receiving the treatment ZDV+ddI or ddI monotherapy as considered in \cite{lu2013variable}.
Treatment indicator $W = 0$ denotes the treatment ddI with 561 subjects, and $W = 1$ denotes the treatment ZDV+ddI with 522 subjects.
Though ACTG175 is a randomized study, there seem to be some selection effects in the subsets used here. For example, for covariate race equals to 1, there are 138 receiving ZDV+ddI and 173 receiving ddI.
A binomial test with null probability 0.5 gives p-value 0.05. For this reason, we analyze the study as an observational rather than randomized study.

Here we are interested in the causal effect between ZDV+ddI and ddI on survival time of HIV-infected patients. 12 selected baseline covariates were studied in \cite{tsiatis2008covariate,zhang2008improving,lu2013variable,fan2017concordance}. There are 5 continuous covariates: age (year), weight (kg), Karnofsky score (scale of 0-100), CD4 count (cells/$\text{mm}^3$) at baseline, CD8 count (cells/$\text{mm}^3$) at baseline. There are 7 binary variables: gender (male = 1, female = 0), homosexual activity (yes = 1, no = 0), race (non-white = 1, white = 0), symptomatic status (symptomatic = 1, asymptomatic = 0), history of intravenous drug use (yes = 1, no = 0), hemophilia (yes = 1, no = 0), and  antiretroviral history (experienced = 1, naive = 0).
As the outcome considered here is the survival time, we also include CD4 count (cells/$\text{mm}^3$) at $20\pm5$ weeks and CD8 count (cells/$\text{mm}^3$) at $20\pm5$ weeks as covariates, in addition to the 12 covariates described above.

\begin{figure}[t]
\centering
\includegraphics[width=0.65\textwidth]{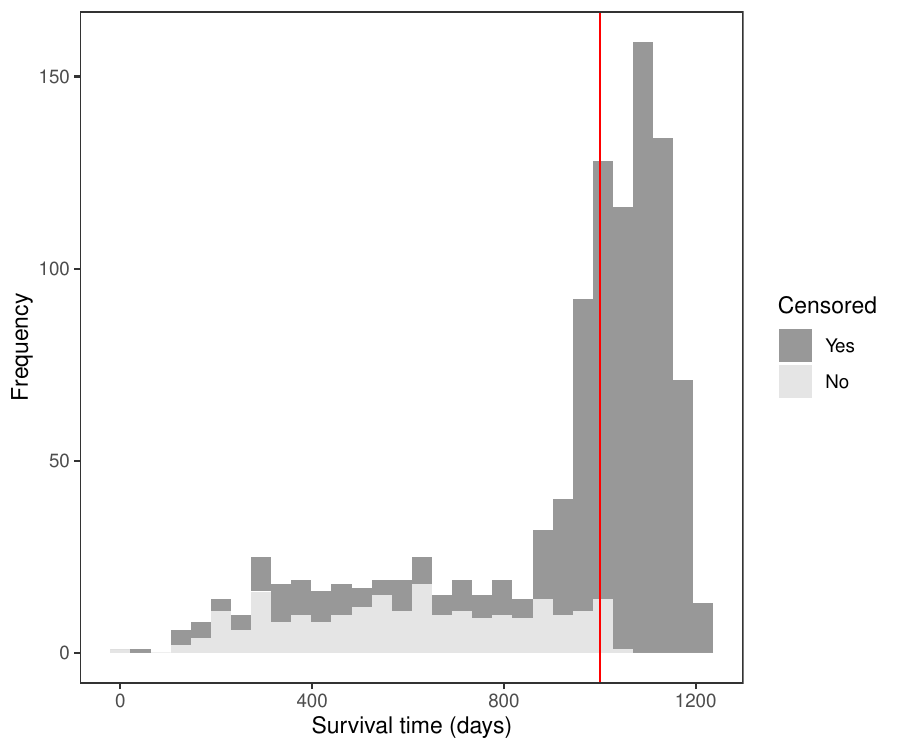}
\caption{Histogram of survival time for censored/non-censored subjects. The solid red line is the suggested truncation time $h = 1000$. The data is from the AIDS Clinical
Trials Group Protocol 175 (ACTG175) \citep{hammer1996trial} with control set to the {\it didanosine} treatment with 561 subjects and the treatment group to {\it zidovudine+didanosine} with 522 subjects.  \label{HIVhistogram}}
\end{figure}

We applied the proposed causal survival forest to this dataset. We used the default tuning parameters, described in Section \ref{sec:simulations}, with the exception of the number of trees which is set to $B=10000$ for computing confidence intervals.
As mentioned in Section \ref{sec:csf}, Assumption \ref{assu:positivity} warrants some extra consideration. The high follow-up time in this study requires care in focusing on an appropriate estimand, namely a suitable $h$ for $T_i \land h$. This study includes a large amount of near end-time censored subjects observed up to almost 6 months after the last failure, which occurs at around 3 years. For this reason, we truncate the survival time right before 3 years, setting $h = 1000$. This assures us that the estimated censoring probabilities $\hat{S}_{W_i}^C\p{U_i\land h\cond X_i}$ all lie in a reasonable range and suggest that we are in a regime where this identifying assumption holds. Figure \ref{HIVhistogram} shows a histogram depicting the issue. In the raw data, the observations with the largest values of $T_i$ are all censored, and so moments of $T_i$ are not identified. However, given a focus on restricted survival time with a judicious choice of $h$,  we get to re-code all (censored or uncensored) observations with $U_i > h$ as uncensored observations with $T_i = h$, thus eliminating the positivity problem.

\begin{figure}[t]
\centering
\includegraphics[width=0.65\textwidth]{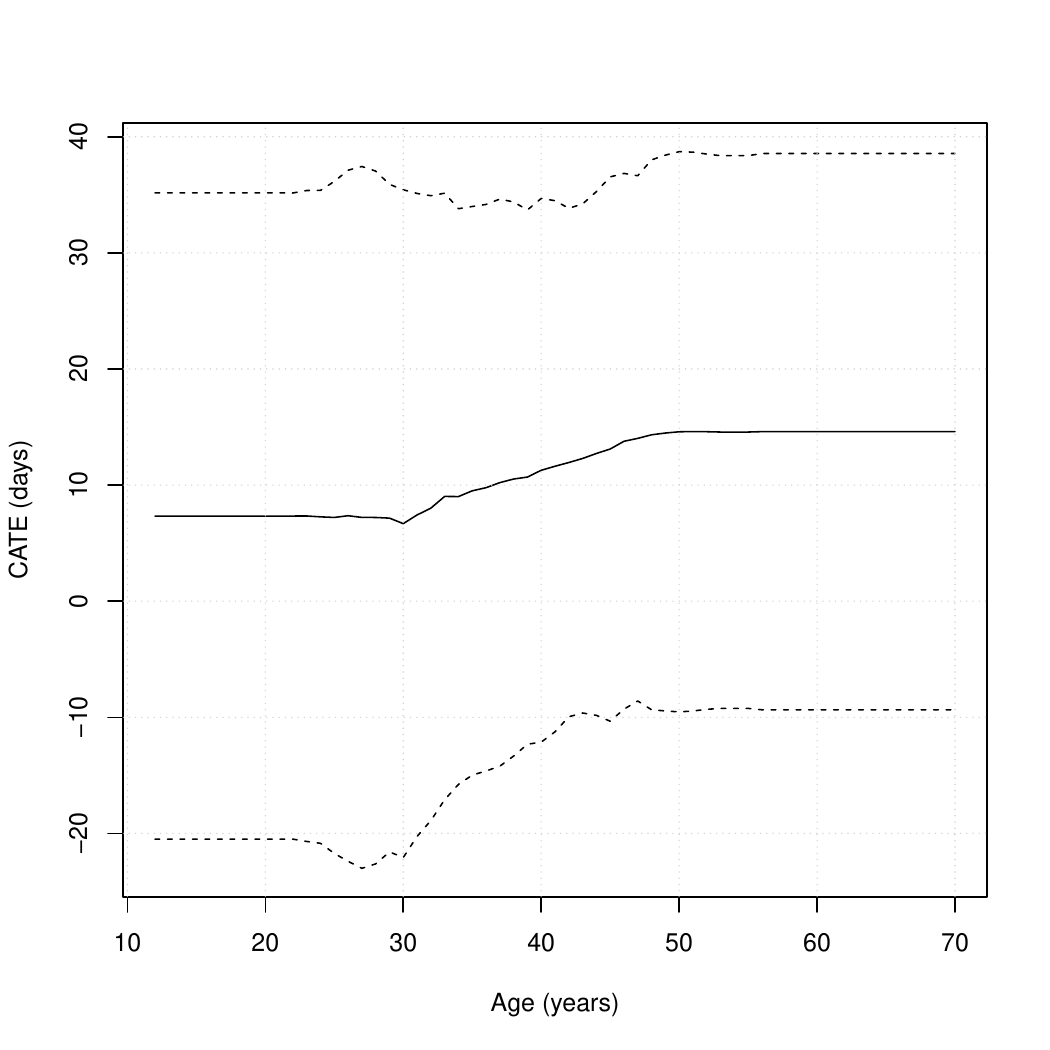}
\caption{Estimated CATEs (survival time in days) versus age (in years) and 95 \% confidence bars (dash lines), with all other covariates set to their median value. CATEs are estimated using a causal survival forest with default options (the number of trees is set to 10 000 for confidence intervals) and the survival time threshold at $1000$ days. The data is from AIDS Clinical
Trials Group Protocol 175 (ACTG175) \citep{hammer1996trial} with control set to the {\it didanosine} treatment with 561 subjects and the treatment group to {\it zidovudine+didanosine} with 522 subjects. 14 covariates are included in the analysis, 5 continuous: age (year), weight (kg), Karnofsky score (scale of 0-100), CD4 count (cells/$\text{mm}^3$) at baseline, and at $20\pm5$ weeks, CD8 count (cells/$\text{mm}^3$) at baseline, and at $20\pm5$ weeks, and 7 binary: gender (male = 1, female = 0), homosexual activity (yes = 1, no = 0), race (non-white = 1, white = 0), symptomatic status (symptomatic = 1, asymptomatic = 0), history of intravenous drug use (yes = 1, no = 0), hemophilia (yes = 1, no = 0), and antiretroviral history (experienced = 1, naive = 0). \label{HIV}}
\end{figure}

Following \cite{lu2013variable} and \cite{fan2017concordance} we consider what role age serves in treatment efficacy. Figure~\ref{HIV} shows the estimated CATEs against age with all other covariates set to their median value, and as the plot suggests, age appears to have a positive effect, with older patients benefiting more from the treatment ZDV+ddI.
Table~\ref{tab:HIVsample} shows point estimates and standard errors from a random sample of patients.
We also consider the best linear projection proposed in Section~\ref{sec:BLP}, in particular, we regress the obtained doubly robust scores on all covariates, and age only. The results in Table~\ref{tab:HIVblp} suggest that we should exercise caution in interpreting heterogeneity in $\hat \tau(\cdot)$, as none of the coefficients in the considered projections are significantly from zero.

\begin{table}[t]
\begin{center}
\begin{tabular}{rrrrllll}
CATE & se(CATE) & Hemophilia & Gender & Homosexual activity & Antiretroviral history \\
\hline
-7.90 & 10.45 & No & Male & Yes & Experienced \\
  -6.90 & 13.89 & No & Male & Yes & Naive \\
  -4.38 & 17.06 & No & Male & Yes & Naive \\
  0.43 & 17.28 & No & Male & No & Naive \\
  7.26 & 19.17 & No & Male & No & Naive \\
  8.20 & 11.45 & No & Male & Yes & Naive \\
  12.80 & 13.82 & No & Female & Yes & Naive \\
  13.42 & 14.16 & No & Male & Yes & Naive \\
  14.95 & 14.13 & No & Male & Yes & Naive \\
  19.99 & 43.53 & Yes & Male & No & Experienced \\
\end{tabular}

\caption{CATE estimates and standard errors from a random sample of 10 individuals as well as patient characteristics corresponding to the four covariates with the highest split frequency obtained from fitting a causal survival forest with default options (survival time threshold at $1000$ days) on data from AIDS Clinical Trials Group Protocol 175. The data is from \citet{hammer1996trial} with control set to the {\it didanosine} treatment with 561 subjects and the treatment group to {\it zidovudine+didanosine} with 522 subjects.  \label{tab:HIVsample}}
\end{center}
\end{table}

\begin{table}[t]
\begin{center}
\begin{tabular}{l D{)}{)}{15)1} D{)}{)}{15)1} }
\hline
 & \multicolumn{1}{c}{All covariates} & \multicolumn{1}{c}{Age only} \\
\hline
Constant               & -151.65 \; (240.10) & -25.85 \; (48.49) \\
Age                    & 1.26 \; (1.46)      & 1.01 \; (1.35)    \\
Weight                 & -1.99 \; (1.02)     &                   \\
Karnofsky score        & 2.75 \; (2.13)      &                   \\
CD4 count              & -0.06 \; (0.13)     &                   \\
CD8 count              & 0.05 \; (0.05)      &                   \\
Gender                 & 44.15 \; (41.90)    &                   \\
Homosexual activity    & -33.08 \; (34.47)   &                   \\
Race                   & 31.61 \; (25.61)    &                   \\
Symptomatic status     & -10.08 \; (35.40)   &                   \\
Intravenous drug use   & 48.36 \; (31.94)    &                   \\
Hemophilia             & -63.46 \; (44.97)   &                   \\
Antiretroviral history & -7.57 \; (24.11)    &                   \\
CD4 count 20+/-5 weeks & -0.03 \; (0.11)     &                   \\
CD8 count 20+/-5 weeks & -0.04 \; (0.04)     &                   \\
\hline
\multicolumn{3}{l}{$HC_3$ \citep{mackinnon1985some} standard errors in parentheses.}
\end{tabular}
\caption{Best linear projection $\hat \tau_{DR}(x) =\hat \beta_0 + A \hat \beta$ on doubly robust estimates obtained from fitting a causal survival forest with default options (survival time threshold at $1000$ days) on data from AIDS Clinical Trials Group Protocol 175. The data is from \citet{hammer1996trial} with control set to the {\it didanosine} treatment with 561 subjects and the treatment group to {\it zidovudine+didanosine} with 522 subjects. 14 covariates are included in the analysis, 5 continuous: age (year), weight (kg), Karnofsky score (scale of 0-100), CD4 count (cells/$\text{mm}^3$) at baseline, and at $20\pm5$ weeks, CD8 count (cells/$\text{mm}^3$) at baseline, and at $20\pm5$ weeks, and 7 binary: gender (male = 1, female = 0), homosexual activity (yes = 1, no = 0), race (non-white = 1, white = 0), symptomatic status (symptomatic = 1, asymptomatic = 0), history of intravenous drug use (yes = 1, no = 0), hemophilia (yes = 1, no = 0), and antiretroviral history (experienced = 1, naive = 0). \label{tab:HIVblp}}
\end{center}
\end{table}

\section*{Acknowledgement}

We are thankful to the three referees, associate editor, and editor for
helpful comments which led to an improved manuscript.
We are grateful to Susan Athey, Scott Fleming, Vitor Hadad, David Hirshberg, Ayush Kanodia, Julie Tibshirani, Yizhe Xu, and Steve Yadlowsky for helpful conversations and suggestions.
We also particularly thank Julie Tibshirani for performing a code review of our implementation and helping us merge it into \texttt{grf}, and
David Hirshberg for contributing to \texttt{grf} the implementation of sample weighting used in our IPCW approach.

\clearpage
\bibliographystyle{plainnat}
\bibliography{causal,survtrees,survtrees2}

\newpage
\appendix

\begin{center}
\huge Appendix
\end{center}

\section{Proof of Proposition~\ref{prop:DR}}\label{sec:appA}

To simplify the notation, we write $T\land h$ as $\tilde T$ and $U\land h$ as $\tilde U$ in this section. We first start with a proof
in the context of estimating $\mu=\mu(x) = \EE{y(T)}$ by the estimating equation~\eqref{eq:SurvExp},
which might also be of independent interest. Namely, we show that
\begin{align}\label{eq:mu}
\hat \mu-\tilde \mu =o_p(\max(c_n^2,c_n d_n)),
\end{align}
where $\tilde \mu$ is an oracle estimator for $\mu$.

\begin{proof}[Proof of~\eqref{eq:mu}]
At a high level, cross-fitting uses cross-fold estimation to avoid bias due to overfitting. Recall that the cross-fitting first splits the data (at random) into two halves $I_1$ and $I_2$,
and then uses an estimator
\begin{align*}
\hat \mu = \frac{n_1}{n} \hat \mu^{I_1} + \frac{n_2}{n} \hat \mu^{I_2},
\end{align*}
where $n_1=|I_1|$, $n_2=|I_2|$, and $\hat \mu^{I_1}, \hat \mu^{I_2}$ are estimated using nuisances estimated from samples $I_2, I_1$, respectively. We essentially need to show that
\begin{align*}
\hat \mu^{I_1}-\tilde \mu^{I_1} =o_p(\max(c_n^2,c_n d_n)),
\end{align*}
where $\tilde \mu_{}^{I_1}$ is the oracle estimator obtained by solving
\begin{align*}
   \frac{1}{n_1} \sum_{i:i\in I_1} \frac{\Delta^h_i (y(U_i)-\mu) }{S_{}^C(\tilde  U_i | X_i)} + \frac{(1 - \Delta^h_i) }{S_{}^C(\tilde U_i | X_i)} \EE{ y(T_i)-\mu \cond X_i, \, \tilde  T_i > \tilde U_i}
- \int_0^{\tilde U_i} \frac{\lambda_{}^C(s | X_i)}{ S_{}^C(s | X_i)}  \EE{ y(T_i) -\mu \cond X_i, \, \tilde T_i > s} ds=0.
\end{align*}
Note that we have the following decomposition of $\hat \mu^{I_1} - \tilde \mu^{I_1}$
\begin{align}\label{eq:decom}
  &  \hat \mu^{I_1} - \tilde \mu^{I_1}   \nonumber  \\
    = &  \frac{1}{n_1}\left[ \sum_{i:i \in I_1} \Delta^h_i \frac{y(U_i)}{\hat S^C_{} (\tilde U_i|X_i)} + (1- \Delta^h_i) \frac{\hat Q_{}(\tilde U_i|X_i)}{{\hat S^C_{} (\tilde U_i|X_i)}} - \int_0^{\tilde U_i} \frac{\hat \lambda_{}^C(s|X_i)}{\hat S^C_{} (s|X_i)}\hat Q_{}(s|X_i)ds\right] \nonumber\\
    & - \frac{1}{n_1}\left[ \sum_{i:i \in I_1} \Delta^h_i \frac{y(U_i)}{S^C_{} (\tilde U_i|X_i)} + (1- \Delta^h_i) \frac{ Q_{}(\tilde U_i|X_i)}{{S^C_{} (\tilde U_i|X_i)}} - \int_0^{\tilde U_i} \frac{\lambda_{}^C(s|X_i)}{S^C_{} (s|X_i)}Q_{}(s|X_i)ds\right] \nonumber\\
= & \frac{1}{n_1} \Bigg [ \sum_{i:i \in I_1} \frac{(1 - \Delta^h_i) }{S_{}^C(\tilde U_i | X_i)}(\hat Q(\tilde U_i|X_i) - Q(\tilde U_i|X_i)) -
\int_0^{\tilde U_i} \frac{\lambda_{}^C(s | X_i)}{ S_{}^C(s | X_i)}  (\hat Q(s|X_i) - Q(s|X_i)) ds \nonumber\\
& +  (1 - \Delta^h_i) (\frac{ 1 }{\hat S_{}^C(\tilde U_i | X_i)}  - \frac{ 1 }{S_{}^C(\tilde U_i | X_i)} )(\hat Q(\tilde U_i|X_i) - Q(\tilde U_i|X_i)) \nonumber\\& -
\int_0^{\tilde U_i} (\frac{\hat \lambda_{}^C(s | X_i)}{ \hat S_{}^C(s | X_i)} - \frac{\lambda_{}^C(s | X_i)}{ S_{}^C(s | X_i)}) (\hat Q(s|X_i) - Q(s|X_i)) ds \nonumber\\
& + (1-\Delta^h_i ) (\frac{1 }{\hat S_{}^C(\tilde U_i | X_i)} -\frac{1 }{S_{}^C(\tilde U_i | X_i)}) Q(\tilde U_i|X_i) + \Delta^h_i (\frac{1 }{\hat S_{}^C(\tilde U_i | X_i)} -\frac{1 }{S_{}^C(\tilde U_i | X_i)}) y(U_i) \nonumber\\
& - \int_0^{\tilde U_i} ( \frac{\hat \lambda_{}^C(s | X_i)}{ \hat S_{}^C(s | X_i)} - \frac{\lambda_{}^C(s | X_i)}{ S_{}^C(s | X_i)} ) Q(s|X_i)ds \Bigg ],
\end{align}
where we denote $\hEE{y(T_i) |\tilde  T_i > \tilde U_i,\, \, X_i}$ by $\hat Q(\tilde U_i|X_i)$, and $\EE{ y(T_i) |\tilde  T_i >\tilde  U_i,\, \, X_i}$ by $Q(\tilde U_i|X_i)$. At a high level, this decomposition separates $\hat \mu^{I_1} - \tilde \mu^{I_1} $ to four terms: two mean zero terms and two product terms.

Note that by the double robustness of equation~\eqref{eq:SurvExp} shown in \citet[][Chapter 10.4]{tsiatis2007semiparametric},
\begin{align*}
 \EE{\frac{(1 - \Delta^h_i) }{S_{}^C(\tilde U_i | X_i)}(\hat Q(\tilde U_i|X_i) - Q(\tilde U_i|X_i)) -
\int_0^{\tilde U_i} \frac{\lambda_{}^C(s | X_i)}{ S_{}^C(s | X_i)}  (\hat Q(s|X_i) - Q(s|X_i)) ds}=0.
\end{align*}
Thanks to our cross-fitting construction, the nuisance components can effectively be treated as deterministic. Thus, after conditioning on $I_2$, the summands used to build the following term become mean-zero and independent:
\begin{align}
&  \EE{ \left(\frac{1}{n_1} \sum_{i:i\in I_1} \frac{(1 - \Delta^h_i) }{S_{}^C(\tilde U_i | X_i)}(\hat Q(\tilde U_i|X_i) - Q(\tilde U_i|X_i)) -
\int_0^{\tilde U_i} \frac{\lambda_{}^C(s | X_i)}{ S_{}^C(s | X_i)}  (\hat Q(s|X_i) - Q(s|X_i)) ds \right)^2} \nonumber \\
= & \EE{ \EE{\left(\frac{1}{n_1} \sum_{i:i\in I_1} \frac{(1 - \Delta^h_i) }{S_{}^C(\tilde U_i | X_i)}(\hat Q(\tilde U_i|X_i) - Q(\tilde U_i|X_i)) -
\int_0^{\tilde U_i} \frac{\lambda_{}^C(s | X_i)}{ S_{}^C(s | X_i)}  (\hat Q(s|X_i) - Q(s|X_i)) ds \right)^2 \Bigg|I_2} }  \nonumber \\
= & \EE{ \Var{\left(\frac{1}{n_1} \sum_{i:i\in I_1} \frac{(1 - \Delta^h_i) }{S_{}^C(\tilde U_i | X_i)}(\hat Q(\tilde U_i|X_i) - Q(\tilde U_i|X_i)) -
\int_0^{\tilde U_i} \frac{\lambda_{}^C(s | X_i)}{ S_{}^C(s | X_i)}  (\hat Q(s|X_i) - Q(s|X_i)) ds \right) \Bigg|I_2} }  \nonumber \\
= & \frac{1}{n_1} \EE{ \Var{\left( \frac{(1 - \Delta^h_i) }{S_{}^C(\tilde U_i | X_i)}(\hat Q(\tilde U_i|X_i) - Q(\tilde U_i|X_i)) -
\int_0^{\tilde U_i} \frac{\lambda_{}^C(s | X_i)}{ S_{}^C(s | X_i)}  (\hat Q(s|X_i) - Q(s|X_i)) ds \right) \Bigg|I_2} } \nonumber \\
\leq & \frac{O_p(1)}{n_1}\sup_{x\in \xx,s\leq h} |\hat Q(s|x)-Q(s|x)|^2 = \frac{o_p(1)}{n}. \label{eq:crossfit}
\end{align}

The same logic applies to the term
\begin{align*}
(1-\Delta^h_i ) (\frac{1 }{\hat S_{}^C(\tilde U_i | X_i)} -\frac{1 }{S_{}^C(\tilde U_i | X_i)}) Q(\tilde U_i|X_i) + \Delta^h_i (\frac{1 }{\hat S_{}^C(\tilde U_i | X_i)} -\frac{1 }{S_{}^C(\tilde U_i | X_i)}) y(U_i) \\
 - \int_0^{\tilde U_i} ( \frac{\hat \lambda_{}^C(s | X_i)}{ \hat S_{}^C(s | X_i)} - \frac{\lambda_{}^C(s | X_i)}{ S_{}^C(s | X_i)} ) Q(s|X_i)ds,
\end{align*}
as
\begin{align*}
\mathbb E \bigg((1-\Delta^h_i ) (\frac{1 }{\hat S_{}^C(\tilde U_i | X_i)} -\frac{1 }{S_{}^C(\tilde U_i | X_i)}) Q(\tilde U_i|X_i) + \Delta^h_i (\frac{1 }{\hat S_{}^C(\tilde U_i | X_i)} -\frac{1 }{S_{}^C(\tilde U_i | X_i)}) y(U_i)\\
 - \int_0^{\tilde U_i} ( \frac{\hat \lambda_{}^C(s | X_i)}{ \hat S_{}^C(s | X_i)} - \frac{\lambda_{}^C(s | X_i)}{ S_{}^C(s | X_i)} ) Q(s|X_i)ds \bigg)=0,
\end{align*}
so we have that
\begin{align}
\mathbb E \bigg( \frac{1}{n_1} \sum_{i:i\in I_1} (1-\Delta^h_i ) (\frac{1 }{\hat S_{}^C(\tilde U_i | X_i)} -\frac{1 }{S_{}^C(\tilde U_i | X_i)}) Q(\tilde U_i|X_i) + \Delta^h_i (\frac{1 }{\hat S_{}^C(\tilde U_i | X_i)} -\frac{1 }{S_{}^C(\tilde U_i | X_i)}) y(U_i) \nonumber \\
 - \int_0^{\tilde U_i} ( \frac{\hat \lambda_{}^C(s | X_i)}{ \hat S_{}^C(s | X_i)} - \frac{\lambda_{}^C(s | X_i)}{ S_{}^C(s | X_i)} ) Q(s|X_i)ds \bigg) \leq \frac{o_p(1)}{n}.
 \label{eq:crossfit2}
\end{align}

In addition, by Cauchy-Schwarz inequality,
\begin{align}
& \frac{1}{n_1}\sum_{i:i\in I_1} \bigg( (1 - \Delta^h_i) (\frac{ 1 }{\hat S_{}^C(\tilde U_i | X_i)}  - \frac{ 1 }{S_{}^C(\tilde U_i | X_i)} )(\hat Q(\tilde U_i|X_i) - Q(\tilde U_i|X_i)) \nonumber \\ &-
\int_0^{\tilde U_i} (\frac{\hat \lambda_{}^C(s | X_i)}{ \hat S_{}^C(s | X_i)} - \frac{\lambda_{}^C(s | X_i)}{ S_{}^C(s | X_i)}) (\hat Q(s|X_i) - Q(s|X_i)) ds  \bigg) \nonumber \\
 \leq & \sqrt{ \frac{1}{n_1} \sum_{i:i\in I_1} (1-\Delta^h_i)(\frac{ 1 }{\hat S_{}^C(\tilde U_i | X_i)}  - \frac{ 1 }{S_{}^C(\tilde U_i | X_i)})^2  } \times
 \sqrt{ \frac{1}{n_1} \sum_{i:i\in I_1} (1-\Delta^h_i)(\hat Q(\tilde U_i|X_i) - Q(\tilde U_i|X_i))^2  } \nonumber \\
 & +\int_0^{\tilde U_i} \sqrt{ \frac{1}{n_1} \sum_{i:i\in I_1} (\frac{ \hat \lambda_{}^C(s | X_i) }{\hat S_{}^C(s | X_i)}  - \frac{  \lambda_{}^C(s | X_i) }{S_{}^C(s | X_i)})^2  } \times
 \sqrt{ \frac{1}{n_1} \sum_{i:i\in I_1} (\hat Q(s|X_i) - Q(s|X_i))^2  } ds \nonumber \\
 = & o_p(\max(c_n^2,c_n d_n)).
 \label{eq:c-s}
\end{align}

Therefore, combining equations~\eqref{eq:crossfit}, \eqref{eq:crossfit2}, and \eqref{eq:c-s}, we have that $\hat \mu-\tilde \mu =o_p(\max(c_n^2,c_n d_n))$.
\end{proof}

Now, we turn to estimating $\tau=\tau(x)=\EE{y(T(1))-y(T(0))}$.
\begin{proof}[Proof of Proposition~\ref{prop:DR}]
Using the same notation as the proof of equation~\eqref{eq:mu} and consider an estimator
\begin{align*}
\hat \tau = \frac{n_1}{n} \hat \tau^{I_1} + \frac{n_2}{n} \hat \tau^{I_2},
\end{align*}
where $n_1=|I_1|$, $n_2=|I_2|$, and $\hat \tau^{I_1}, \hat \tau^{I_2}$ are estimated using nuisances estimated from samples $I_2, I_1$, respectively.
Note that
\begin{align*}
   & \hat \tau^{I_1} - \tilde \tau^{I_1}\\
    = & \bigg( \frac{1}{n_1}\sum_{i:i\in I_1} (W_i- \hat e(X_i))^2 \bigg)^{-1} \frac{1}{n_1}\sum_{i:i\in I_1} \bigg(  \frac{\hat \Upsilon_i}{\hat S^C_{W_i} (\tilde U_i|X_i)} - \int_0^{\tilde U_i} \frac{\hat \lambda_{W_i}^C(s|X_i)}{\hat S^C_{W_i} (s|X_i)}(W_i-\hat e(X_i))[\hat Q_{W_i}(s|X_i) -\hat m(X_i) ]ds \bigg)\\&-\bigg(\frac{1}{n_1} \sum_{i:i\in I_1} (W_i- e(X_i))^2  \bigg)^{-1} \frac{1}{n_1} \sum_{i:i\in I_1} \bigg(  \frac{\Upsilon_i}{ S^C_{W_i} (\tilde U_i|X_i)} - \int_0^{\tilde U_i} \frac{ \lambda_{W_i}^C(s|X_i)}{ S^C_{W_i} (s|X_i)}(W_i- e(X_i))[ Q_{W_i}(s|X_i) - m(X_i) ]ds \bigg),
\end{align*}
where $\tilde \tau_{}^{I_1}$ is the oracle estimator,
\begin{align*}
\Upsilon_i&=
\begin{cases}
\{W_i-e(X_i)\}\{y(U_i)-m(X_i)\} & \text{if}\quad  \Delta^h_i=1,\\
\{W_i-e(X_i)\}\{Q_{W_i}(\tilde U_i|X_i)-m(X_i)\} & \text{o.w.}
\end{cases}
\end{align*}
and
\begin{align*}
\hat \Upsilon_i&=
\begin{cases}
\{W_i-\hat e(X_i)\}\{y(U_i)-\hat m(X_i)\} & \text{if}\quad  \Delta^h_i=1,\\
\{W_i-\hat e(X_i)\}\{\hat Q_{W_i}(\tilde U_i|X_i)-\hat m(X_i)\} & \text{o.w.}
\end{cases}
\end{align*}

Denote $\hat \tau^{I_1} - \tilde \tau^{I_1}$ by
\begin{align*}
 \frac{K_1}{J_1}-\frac{K_2}{J_2}.
\end{align*}
Thus, we have
\begin{align}\label{eq:IJ}
& \frac{K_1}{J_1}-\frac{K_2}{J_2}=\frac{K_1+K_2-K_2}{J_1}-\frac{K_2}{J_2} \nonumber \\
= & \frac{K_1-K_2}{J_1}+ K_2(\frac{1}{J_1}-\frac{1}{J_2})= \frac{K_1-K_2}{J_1}+ \frac{K_2}{J_1J_2}(J_2-J_1).
\end{align}
We essentially need to bound $K_1-K_2$, and we have the following decomposition,
\begin{align*}
    K_1 -K_2
    = & \frac{1}{n_1}\bigg[\sum_{i:i \in I_1} \Delta^h_i \frac{\{W_i-\hat e(X_i)\}\{y(U_i)-\hat m(X_i)\}}{\hat S^C_{W_i} (\tilde U_i|X_i)} + (1- \Delta^h_i) \frac{\{W_i-\hat e(X_i)\}\{\hat Q_{W_i}(\tilde U_i|X_i)-\hat m(X_i)\}}{{\hat S^C_{W_i} (\tilde U_i|X_i)}} \\
    & - \int_0^{\tilde U_i} \frac{\hat \lambda_{W_i}^C(s|X_i)}{\hat S^C_{W_i} (s|X_i)}(W_i-\hat e(X_i))[\hat Q_{W_i}(s|X_i) -\hat m(X_i) ]ds\\
                                    & - \Delta^h_i \frac{\{W_i- e(X_i)\}\{y(U_i)- m(X_i)\}}{ S^C_{W_i} (\tilde U_i|X_i)} - (1- \Delta^h_i) \frac{\{W_i- e(X_i)\}\{ Q_{W_i}(\tilde U_i|X_i)- m(X_i)\}}{{ S^C_{W_i} (\tilde U_i|X_i)}}\\
        & + \int_0^{\tilde U_i} \frac{ \lambda_{W_i}^C(s|X_i)}{ S^C_{W_i} (s|X_i)}(W_i- e(X_i))[ Q_{W_i}(s|X_i) - m(X_i) ]ds\bigg]
    \\
        = & \frac{1}{n_1}\bigg[\sum_{i:i \in I_1} \Delta^h_i \frac{\{W_i-\hat e(X_i)\}\{y(U_i)-\hat m(X_i)\}}{\hat S^C_{W_i} (\tilde U_i|X_i)} + (1- \Delta^h_i) \frac{\{W_i-\hat e(X_i)\}\{\hat Q_{W_i}(\tilde U_i|X_i)-\hat m(X_i)\}}{{\hat S^C_{W_i} (\tilde U_i|X_i)}} \\
        & - \int_0^{\tilde U_i} \frac{\hat \lambda_{W_i}^C(s|X_i)}{\hat S^C_{W_i} (s|X_i)}(W_i-\hat e(X_i))[\hat Q_{W_i}(s|X_i) -\hat m(X_i) ]ds
    \\
        & - \Delta^h_i \frac{\{W_i-\hat e(X_i)\}\{y(U_i)-\hat m(X_i)\}}{ S^C_{W_i} (\tilde U_i|X_i)} - (1- \Delta^h_i) \frac{\{W_i-\hat e(X_i)\}\{ Q_{W_i}(\tilde U_i|X_i)-\hat m(X_i)\}}{{ S^C_{W_i} (\tilde U_i|X_i)}}\\
        & + \int_0^{\tilde U_i} \frac{ \lambda_{W_i}^C(s|X_i)}{ S^C_{W_i} (s|X_i)}(W_i-\hat e(X_i))[ Q_{W_i}(s|X_i) -\hat m(X_i) ]ds \\
                & + \Delta^h_i \frac{\{W_i-\hat e(X_i)\}\{y(U_i)-\hat m(X_i)\}}{ S^C_{W_i} (\tilde U_i|X_i)} + (1- \Delta^h_i) \frac{\{W_i-\hat e(X_i)\}\{ Q_{W_i}(\tilde U_i|X_i)-\hat m(X_i)\}}{{ S^C_{W_i} (\tilde U_i|X_i)}}\\
        & - \int_0^{\tilde U_i} \frac{ \lambda_{W_i}^C(s|X_i)}{ S^C_{W_i} (s|X_i)}(W_i-\hat e(X_i))[ Q_{W_i}(s|X_i) -\hat m(X_i) ]ds \\
                        & - \Delta^h_i \frac{\{W_i- e(X_i)\}\{y(U_i)- m(X_i)\}}{ S^C_{W_i} (\tilde U_i|X_i)} - (1- \Delta^h_i) \frac{\{W_i- e(X_i)\}\{ Q_{W_i}(\tilde U_i|X_i)- m(X_i)\}}{{ S^C_{W_i} (\tilde U_i|X_i)}}\\
        & + \int_0^{\tilde U_i} \frac{ \lambda_{W_i}^C(s|X_i)}{ S^C_{W_i} (s|X_i)}(W_i- e(X_i))[ Q_{W_i}(s|X_i) - m(X_i) ]ds \bigg].
\end{align*}
At a high level, this decomposition separates $K_1-K_2$ into two terms: the first term takes $\hat{m}$ and $\hat{e}$ as given, and follow the construction in the survival-related nuisance components \eqref{eq:decom} to bound errors caused by using $\hat{\lambda}^C_w,  \, \hat{S}^C_w, \, \hat{Q}_w$ instead of ${\lambda}^C_w,  \, {S}^C_w, \, {Q}_w$; the second term bounds errors caused by using $\hat{m}$ and $\hat{e}$ instead of ${m}$ and ${e}$.

For the term
\begin{align*}
 &   \Delta^h_i \frac{\{W_i-\hat e(X_i)\}\{y(U_i)-\hat m(X_i)\}}{\hat S^C_{W_i} (\tilde U_i|X_i)} + (1- \Delta^h_i) \frac{\{W_i-\hat e(X_i)\}\{\hat Q_{W_i}(\tilde U_i|X_i)-\hat m(X_i)\}}{{\hat S^C_{W_i} (\tilde U_i|X_i)}} \\
        & - \int_0^{\tilde U_i} \frac{\hat \lambda_{W_i}^C(s|X_i)}{\hat S^C_{W_i} (s|X_i)}(W_i-\hat e(X_i))[\hat Q_{W_i}(s|X_i) -\hat m(X_i) ]ds
    \\
        & - \Delta^h_i \frac{\{W_i-\hat e(X_i)\}\{y(U_i)-\hat m(X_i)\}}{ S^C_{W_i} (U_i|X_i)} - (1- \Delta^h_i) \frac{\{W_i-\hat e(X_i)\}\{ Q_{W_i}(\tilde U_i|X_i)-\hat m(X_i)\}}{{ S^C_{W_i} (\tilde U_i|X_i)}}\\
        & + \int_0^{\tilde U_i} \frac{ \lambda_{W_i}^C(s|X_i)}{ S^C_{W_i} (s|X_i)}(W_i-\hat e(X_i))[ Q_{W_i}(s|X_i) -\hat m(X_i) ]ds \equiv (I),
\end{align*}
the proof follows from the same decomposition of equation~\eqref{eq:decom} with $\tilde U_i$ replaced by $\{W_i-\hat e(X_i)\}\{\tilde U_i-\hat m(X_i)\}$ and $\hat Q(\tilde U_i|X_i)$ replaced by $\{W_i-\hat e(X_i)\}\{\hat Q_{W_i}(\tilde U_i|X_i)-\hat m(X_i)\}$. So we have that
\begin{align}\label{eq:1rate}
  (I) = o_p(c_n^2 + c_nd_n).
\end{align}

For the term
\begin{align*}
                    & \Delta^h_i \frac{\{W_i-\hat e(X_i)\}\{y(U_i)-\hat m(X_i)\}}{ S^C_{W_i} (\tilde U_i|X_i)} + (1- \Delta^h_i) \frac{\{W_i-\hat e(X_i)\}\{ Q_{W_i}(\tilde U_i|X_i)-\hat m(X_i)\}}{{ S^C_{W_i} (\tilde U_i|X_i)}}\\
        & - \int_0^{\tilde U_i} \frac{ \lambda_{W_i}^C(s|X_i)}{ S^C_{W_i} (s|X_i)}(W_i-\hat e(X_i))[ Q_{W_i}(s|X_i) -\hat m(X_i) ]ds \\
                        & - \Delta^h_i \frac{\{W_i- e(X_i)\}\{y(U_i)- m(X_i)\}}{ S^C_{W_i} (\tilde U_i|X_i)} - (1- \Delta^h_i) \frac{\{W_i- e(X_i)\}\{ Q_{W_i}(\tilde U_i|X_i)- m(X_i)\}}{{ S^C_{W_i} (\tilde U_i|X_i)}}\\
        & + \int_0^{\tilde U_i} \frac{ \lambda_{W_i}^C(s|X_i)}{ S^C_{W_i} (s|X_i)}(W_i- e(X_i))[ Q_{W_i}(s|X_i) - m(X_i) ]ds \equiv (II),
\end{align*}
we have the following decomposition of $(II)$,
\begin{align*}
                    & \frac{\Delta^h_i}{S^C_{W_i} (\tilde U_i|X_i)} [ (e(X_i)-\hat e(X_i))(m(X_i)-\hat m(X_i)) + (e(X_i)-\hat e(X_i))(y(U_i)-m(X_i)) + (W_i-e(X_i))(m(X_i)-\hat m(X_i))] \\
                   & + \frac{(1- \Delta^h_i)} { S^C_{W_i} (\tilde U_i|X_i)} [ (e(X_i)-\hat e(X_i))(m(X_i)-\hat m(X_i)) + (e(X_i)-\hat e(X_i))(Q_{W_i}(\tilde U_i|X_i)-m(X_i))\\& + (W_i-e(X_i))(m(X_i)-\hat m(X_i))] \\
        & - \int_0^{\tilde U_i} \frac{ \lambda_{W_i}^C(s|X_i)}{ S^C_{W_i} (s|X_i)}[ (e(X_i)-\hat e(X_i))(m(X_i)-\hat m(X_i)) + (e(X_i)-\hat e(X_i))(Q_{W_i}(s|X_i)-m(X_i))\\& + (W_i-e(X_i))(m(X_i)-\hat m(X_i))] ds.
\end{align*}
Note that
\begin{align*}
&\EE{\frac{\Delta^h_i}{S^C_{W_i} (\tilde U_i|X_i)}(e(X_i)-\hat e(X_i))(y(U_i)-m(X_i))}=0,\\
&\EE{\frac{1}{S^C_{W_i} (\tilde U_i|X_i)}(W_i- e(X_i))(m(X_i)-\hat m(X_i))-\int_0^{\tilde U_i} \frac{ \lambda_{W_i}^C(s|X_i)}{ S^C_{W_i} (s|X_i)} (W_i-e(X_i))(m(X_i)-\hat m(X_i)) ds}=0,
\end{align*}
and
{\small \begin{align*}
\EE{\frac{(1- \Delta^h_i)} { S^C_{W_i} (\tilde U_i|X_i)}  (e(X_i)-\hat e(X_i))(Q_{W_i}(\tilde U_i|X_i)-m(X_i)) - \int_0^{\tilde U_i} \frac{ \lambda_{W_i}^C(s|X_i)}{ S^C_{W_i} (s|X_i)} (e(X_i)-\hat e(X_i))(Q_{W_i}(s|X_i)-m(X_i)) ds} =0.
\end{align*}
} Again, by the law of iterated expectations similar to that of \eqref{eq:crossfit} using cross-fitting technique and Cauchy-Schwarz inequality in \eqref{eq:c-s}, we have that
\begin{align}\label{eq:2rate}
  (II) = o_p(b_n c_n).
\end{align}
Combining equations~\eqref{eq:1rate} and \eqref{eq:2rate}, we have that
$K_1-K_2=o_p(c_nd_n+c_n^2+b_nc_n)$.
Further combining the rates of $K_1-K_2$ and $J_2-J_1$ using equation~\eqref{eq:IJ}, we have that \begin{align*}
 \hat \tau^{I_1}- \tilde \tau_{}^{I_1}
= o_p( \max( (c_n+d_n)c_n,b_n c_n,b_n^2) ),
\end{align*}
and therefore  $\hat \tau- \tilde \tau
= o_p( \max( (c_n+d_n)c_n,b_n c_n,b_n^2) )$.
Recall that $\tilde \tau$ is an i.i.d. average, so we immediately have that by using cross-fitting, we transform any $n^{1/4}$ consistent machine learning method into an efficient estimator of $\tau = \EE{ y(T(1))-y(T(0))}$.

\end{proof}

\section{Proof of Lemma~\ref{lemma:couple}}

\begin{proof}
Note that
%{\footnotesize
\begin{align*}
   & \hat \tau(x) - \tilde \tau(x)\\
    = & \bigg( \frac{1}{l}\sum_{i=1}^l (W_i- \hat e(X_i))^2 \bigg)^{-1} \frac{1}{l}\sum_{i=1}^l \bigg(  \frac{\hat \Upsilon_i}{\hat S^C_{W_i} (U_i\land h|X_i)} - \int_0^{U_i\land h} \frac{\hat \lambda_{W_i}^C(t|X_i)}{\hat S^C_{W_i} (t|X_i)}(W_i-\hat e(X_i))[\hat Q_{W_i}(t|X_i) -\hat m(X_i) ]dt \bigg)\\&-\bigg(\frac{1}{l} \sum_{i=1}^l (W_i- e(X_i))^2  \bigg)^{-1} \frac{1}{l} \sum_{i=1}^l \bigg(  \frac{\Upsilon_i}{ S^C_{W_i} (U_i\land h|X_i)} - \int_0^{U_i\land h} \frac{ \lambda_{W_i}^C(t|X_i)}{ S^C_{W_i} (t|X_i)}(W_i- e(X_i))[ Q_{W_i}(t|X_i) - m(X_i) ]dt \bigg),
\end{align*}
%}
where $\Upsilon_i, \hat \Upsilon_i$ are defined in Section~\ref{sec:appA} of the Appendix, $l$ is the number of observations falling into the same terminal node as $x$, and the weights $\alpha_i(x)$ are absorbed in $l$.
Following the same proof of Proposition~\ref{prop:DR} except the Cauchy-Schwarz inequality such as equation~\eqref{eq:c-s} becomes
\begin{align*}
& \frac{1}{l}\sum_{i=1}^l \bigg( (1 - \Delta^h_i) (\frac{ 1 }{\hat S_{}^C(\tilde U_i | X_i)}  - \frac{ 1 }{S_{}^C(\tilde U_i | X_i)} )(\hat Q(\tilde U_i|X_i) - Q(\tilde U_i|X_i)) \nonumber \\ &-
\int_0^{\tilde U_i} (\frac{\hat \lambda_{}^C(s | X_i)}{ \hat S_{}^C(s | X_i)} - \frac{\lambda_{}^C(s | X_i)}{ S_{}^C(s | X_i)}) (\hat Q(s|X_i) - Q(s|X_i)) ds  \bigg) \nonumber \\
  \leq & \sqrt{ \frac{1}{l} \sum_{i=1}^l (1-\Delta^h_i)\sup_{x\in \XX}(\frac{ 1 }{\hat S_{}^C(\tilde U_i | x)}  - \frac{ 1 }{S_{}^C(\tilde U_i | x)})^2  } \times
 \sqrt{ \frac{1}{l} \sum_{i=1}^l (1-\Delta^h_i)\sup_{x\in \XX}(\hat Q(\tilde U_i|x) - Q(\tilde U_i|x))^2  } \nonumber \\
 & +\int_0^{\tilde U_i} \sqrt{ \frac{1}{l} \sum_{i=1}^l \sup_{x\in \XX}(\frac{ \hat \lambda_{}^C(s | x) }{\hat S_{}^C(s | x)}  - \frac{  \lambda_{}^C(s | x) }{S_{}^C(s | x)})^2  } \times
 \sqrt{ \frac{1}{l} \sum_{i=1}^l \sup_{x\in \XX}(\hat Q(s|x) - Q(s|x))^2  } ds \nonumber \\
 = & o_p(\max(c_n^2,c_n d_n)),
\end{align*}
we have that
\begin{align*}
\hat \tau(x)- \tilde \tau(x) = o_p(\max((c_n+d_n)c_n,b_n c_n,b^2_n)),
\end{align*}
which completes the proof.
\end{proof}

\section{Proof of Theorem~\ref{theorem:main}}

\begin{proof}

Given the set of forest weights $\alpha_i(x)$ used to define the generalized random forest estimation $\tilde \tau(x)$ with unknown true nuisance parameters, we have the following linear approximation
\begin{equation*}
\tilde \tau^*(x)=\tau(x)+\sum_{i=1}^n \alpha_i(x) \rho_i^*(x),
\end{equation*}
where $\rho_j^*(x)$ denotes the influence function of the $j$-th observation with respect to the true parameter value $\tau(x)$, and $\tilde \tau^*(x)$ is a pseudo-forest output with weights $\alpha_i(x)$ and outcomes $\tau(x)+\rho^*_i(x)$.

Note that Assumptions 2-6  in \cite{wager2019grf} hold immediately from the definition of the estimating equation $\psi_{\tau(x)}$. In particular, $\psi_{\tau(x)}$ is Lipschitz continuous in terms of $\tau(x)$ for their Assumption 4; The solution of $\sum_{i=1}^n \alpha_i \psi^i_{\tau(x)}=0$ always exists for their Assumption 5.
By the results shown in \cite{wager2018estimation}, there exists a sequence $\sigma_n(x)$ for which $$[\tilde \tau^*(x) -\tau(x)]/\sigma_n(x) \rightarrow N(0,1),$$ where $\sigma^2_n(x) = polylog(n/\ell)^{-1}\ell/n$ and $polylog(n/\ell)$ is a function that is bounded away from 0 and increases at most polynomially with the log-inverse sampling ratio $\log (n/\ell)$.

Furthermore, by Lemma~4 in \cite{wager2019grf},
\begin{align*}
({n}/{\ell})^{1/2} [\tilde \tau(x)-\tilde \tau^*(x)] = O_p(\max( \ell^{-\frac{\pi\log((1-\nu)^{-1})}{2\log(\nu^{-1})}}, (\frac{\ell}{n})^{1/6})).
\end{align*}
Following Lemma~\ref{lemma:couple}, as long as $o_p(\max((c_n+d_n)c_n,b_n c_n,b^2_n))$ goes faster than $polylog(n/\ell)^{-1/2}(\ell/n)^{1/2}$, we have
$$[\hat \tau(x) -\tau(x)]/\sigma_n(x) \rightarrow N(0,1).$$

\end{proof}

\section{Figures}\label{sec:simfigures}

\begin{figure}[H]
\centering
\includegraphics[width=0.95\textwidth]{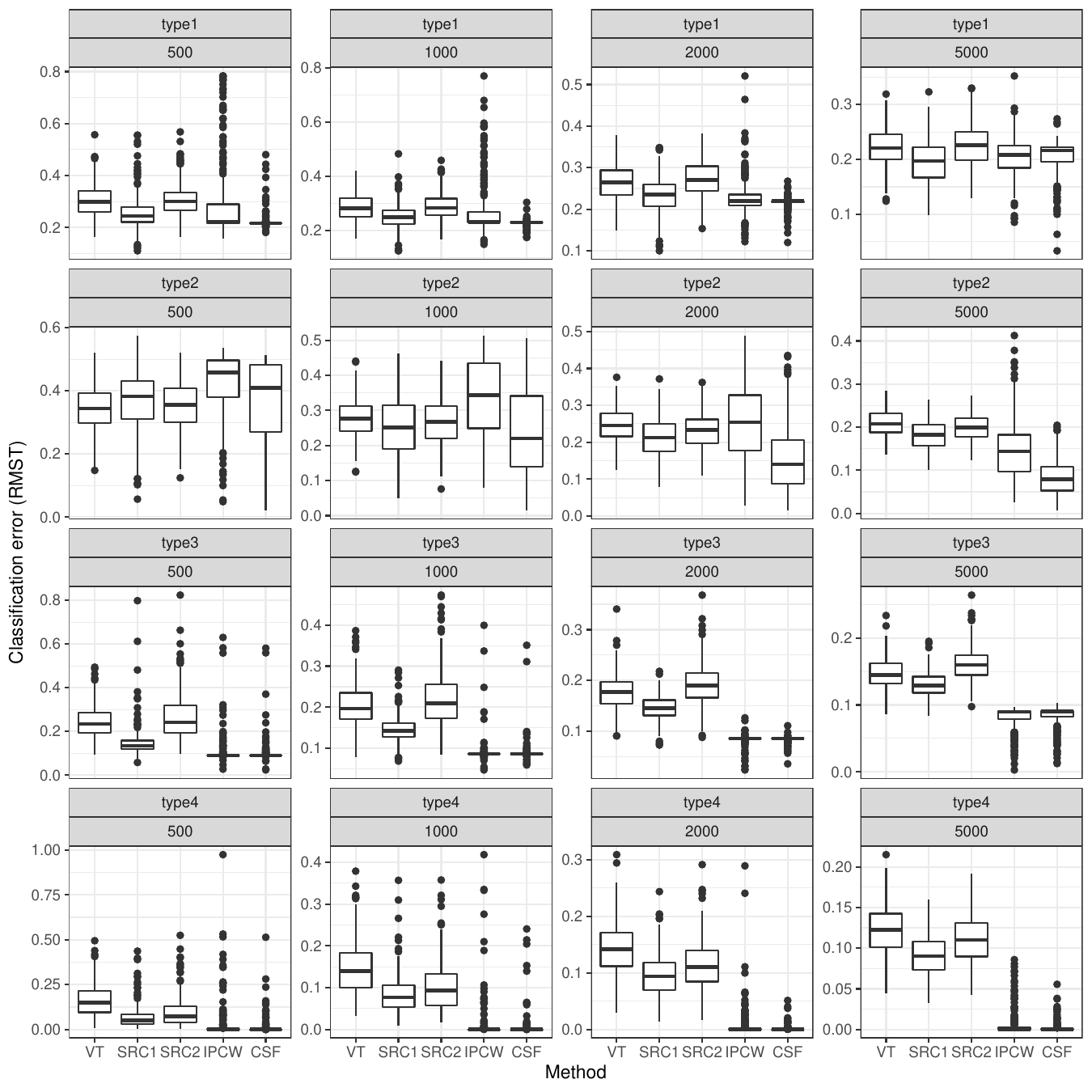}
\caption{Classification error of different methods with RMST estimand. From left to right, top to bottom, the plots correspond to Scenarios 1-4, respectively.  ``VT'' denotes the virtual twin method with random survival forests; ``SRC1'' denotes random survival forests using covariates $(X, W)$; ``SRC2'' denotes random survival forests using covariates $(X,W,XW)$. ``IPCW'' denotes a causal forest with Inverse Probability of Censoring Weighting; ``CSF'' denotes causal survival forest. Training size is (500, 1000, 2000 500), the number of covariates 15, the size of the test set 2000, and number of repetitions 250. \label{simCLF_RMST}}
\end{figure}

\begin{figure}[H]
\centering
\includegraphics[width=0.95\textwidth]{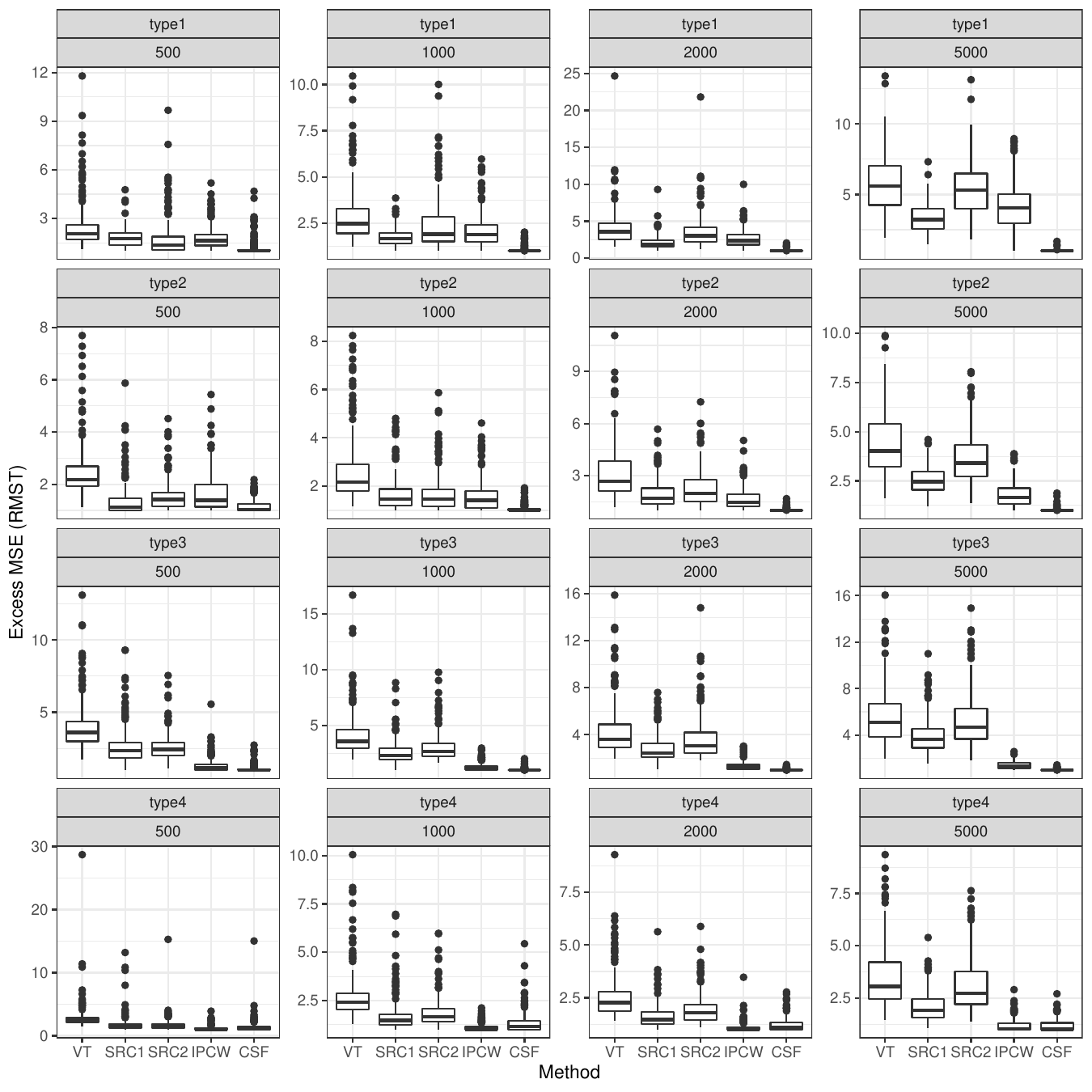}
\caption{Excess MSE (defined as $\frac{1}{B} \sum_{b=1}^{B} MSE_b(method) / min(MSE_b(m)) : m \in \{all \quad methods\}$) for different methods with RMST estimand in the four scenarios. From left to right, top to bottom, the plots correspond to Scenarios 1-4, respectively.  ``VT'' denotes the virtual twin method with random survival forests; ``SRC1'' denotes random survival forests using covariates $(X, W)$; ``SRC2'' denotes random survival forests using covariates $(X,W,XW)$. ``IPCW'' denotes a causal forest with Inverse Probability of Censoring Weighting; ``CSF'' denotes causal survival forest. Training size is (500, 1000, 2000, 5000), the number of covariates 15, the size of the test set 2000, and the number of repetitions 250. \label{simMSE_RMST}}
\end{figure}

\begin{figure}[H]
\centering
\includegraphics[width=0.95\textwidth]{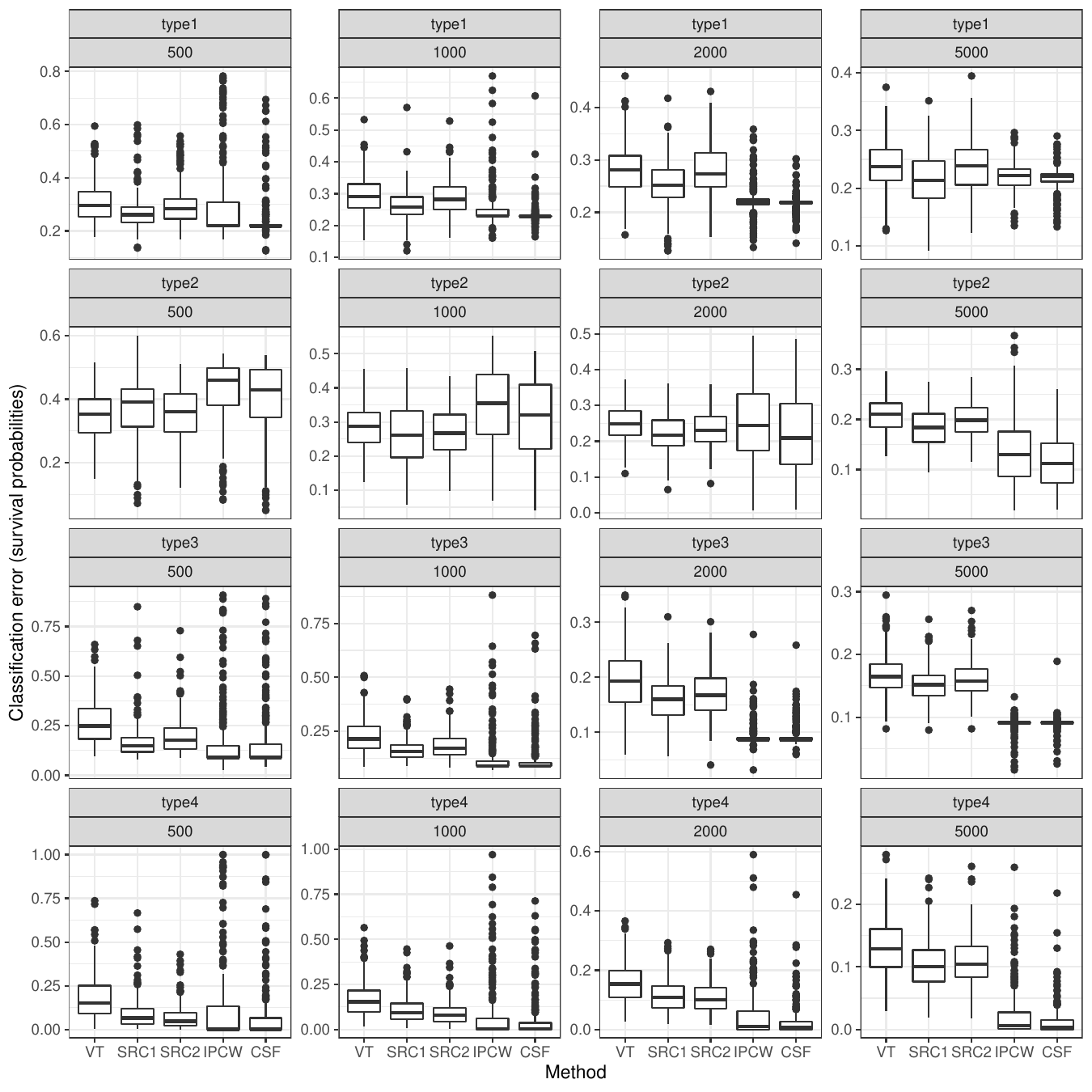}
\caption{Classification error of different methods with survival probability estimand. From left to right, top to bottom, the plots correspond to Scenarios 1-4, respectively.  ``VT'' denotes the virtual twin method with random survival forests; ``SRC1'' denotes random survival forests using covariates $(X, W)$; ``SRC2'' denotes random survival forests using covariates $(X,W,XW)$. ``IPCW'' denotes a causal forest with Inverse Probability of Censoring Weighting; ``CSF'' denotes causal survival forest. Training size is (500, 1000, 2000 500), the number of covariates 15, the size of the test set 2000, and number of repetitions 250. \label{simuCLF_SP}}
\end{figure}

\begin{figure}[H]
\centering
\includegraphics[width=0.95\textwidth]{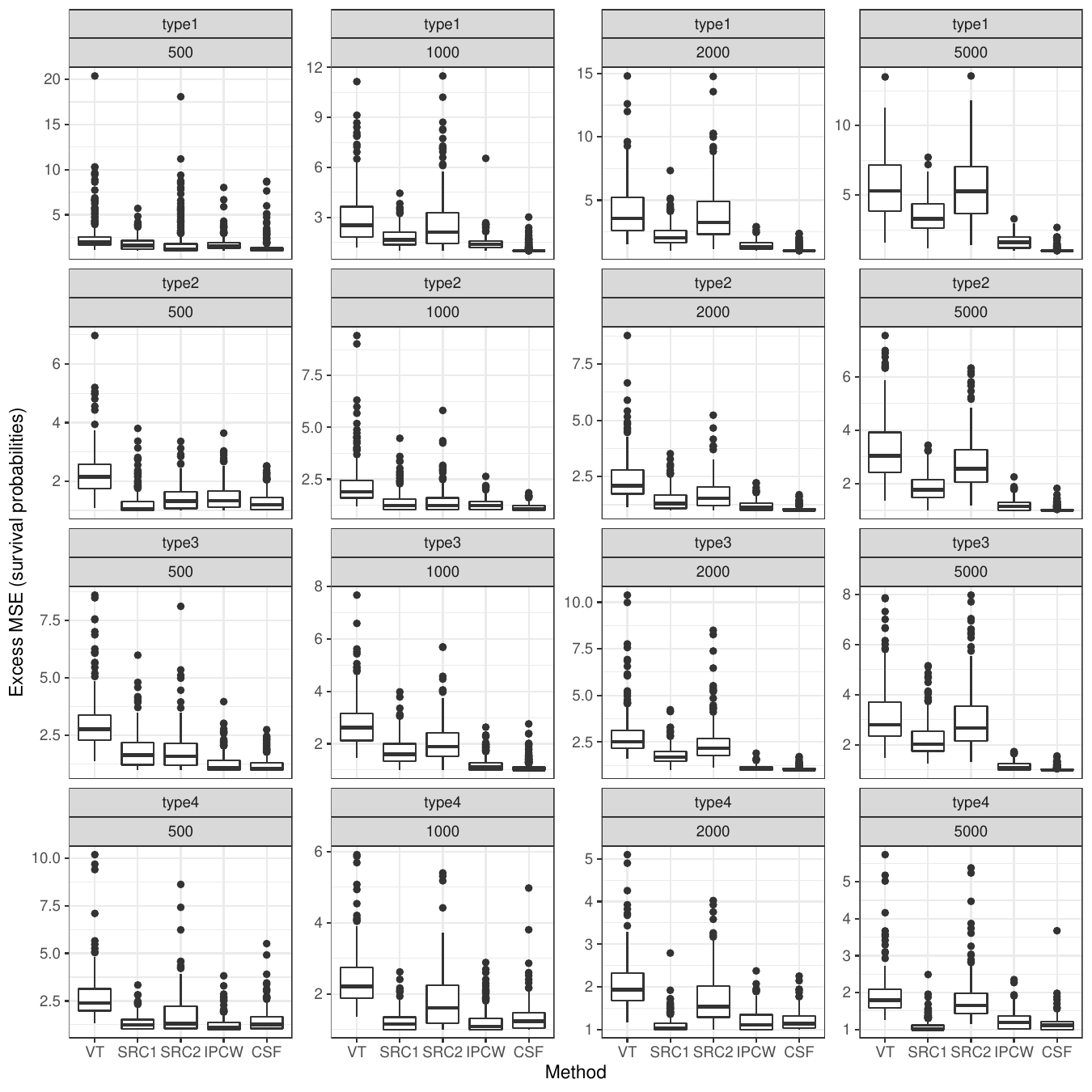}
\caption{Excess MSE (defined as $\frac{1}{B} \sum_{b=1}^{B} MSE_b(method) / min(MSE_b(m)) : m \in \{all \quad methods\}$) for different methods with survival probability estimand in the four scenarios. From left to right, top to bottom, the plots correspond to Scenarios 1-4, respectively.  ``VT'' denotes the virtual twin method with random survival forests; ``SRC1'' denotes random survival forests using covariates $(X, W)$; ``SRC2'' denotes random survival forests using covariates $(X,W,XW)$. ``IPCW'' denotes a causal forest with Inverse Probability of Censoring Weighting; ``CSF'' denotes causal survival forest. Training size is (500, 1000, 2000, 5000), the number of covariates 15, the size of the test set 2000, and the number of repetitions 250. \label{simMSE_SP}}
\end{figure}

\end{document}